\documentclass[apj,revtex4]{emulateapj}
\usepackage{graphicx}

\shorttitle{SN 2012cg}
\shortauthors{Marion et al.}

\newcommand{\cii}{\ion{C}{2}}

\newcommand{\si}{\ion{Si}{2}}

\newcommand{\ca}{\ion{Ca}{2}}
\newcommand{\nai}{\ion{Na}{1}}

\newcommand{\kms}{km s$^{-1}$}

\newcommand{\bmax}{$t(B_{max})$}
\newcommand{\wl}{$\lambda$}

\begin{document} 

\title{SN~2012\lowercase{cg}: Evidence for Interaction Between a Normal Type I\lowercase{a} Supernova\\
        and a Non-Degenerate Binary Companion}

\author{
G.~H.~Marion\altaffilmark{1,2},
Peter~J.~Brown\altaffilmark{3},
Jozsef~Vink\'o\altaffilmark{1,4},
Jeffrey~M.~Silverman\altaffilmark{1,5},
David~J.~Sand\altaffilmark{6},
Peter~Challis\altaffilmark{2},
Robert~P.~Kirshner\altaffilmark{2},
J.~Craig~Wheeler\altaffilmark{1},
Perry~Berlind\altaffilmark{2},
Warren~R.~Brown\altaffilmark{2},
Michael~L.~Calkins\altaffilmark{2},
Yssavo~Camacho\altaffilmark{7,8},
Govinda~Dhungana\altaffilmark{9},
Ryan~J.~Foley\altaffilmark{10,11},
Andrew~S.~Friedman\altaffilmark{12,2},
Melissa~L.~Graham\altaffilmark{13},
D.~Andrew~Howell\altaffilmark{14,15},
Eric~Y.~Hsiao\altaffilmark{16,17},
Jonathan~M.~Irwin\altaffilmark{2}, 
Saurabh~W.~Jha\altaffilmark{7},
Robert~Kehoe\altaffilmark{9},
Lucas~M.~Macri\altaffilmark{3},
Keiichi~Maeda\altaffilmark{18,19},
Kaisey~Mandel\altaffilmark{2},
Curtis~McCully\altaffilmark{14},
Viraj~Pandya\altaffilmark{7,20},
Kenneth~J.~Rines\altaffilmark{21},
Steven~Wilhelmy\altaffilmark{21}
 and Weikang~Zheng\altaffilmark{13}
}

\altaffiltext{1}{University of Texas at Austin, 1 University Station C1400, Austin, TX, 78712-0259, USA}
\altaffiltext{2}{Harvard-Smithsonian Center for Astrophysics, 60 Garden St., Cambridge, MA 02138, USA; \email{ghmarion@gmail.com}}
\altaffiltext{3}{George P. and Cynthia Woods Mitchell Institute for Fundamental Physics \& Astronomy, Texas A. \& M. University, Department of Physics and Astronomy, 4242 TAMU, College Station, TX 77843, USA}
\altaffiltext{4}{Department of Optics and Quantum Electronics, University of Szeged, Domter 9, 6720, Szeged, Hungary}
\altaffiltext{5}{NSF Astronomy and Astrophysics Postdoctoral Fellow}
\altaffiltext{6}{Physics Department, Texas Tech University, Lubbock, TX , 79409, USA}
\altaffiltext{7}{Department of Physics and Astronomy, Rutgers the State University of New Jersey, 136 Frelinghuysen Road, Piscataway, NJ 08854 USA}
\altaffiltext{8}{Department of Physics, Lehigh University, 16 Memorial Drive East, Bethlehem, Pennsylvania 18015, USA}
\altaffiltext{9}{Department of Physics, Southern Methodist University, Dallas, TX 75275, USA}
\altaffiltext{10}{Astronomy Department, University of Illinois at Urbana-Champaign,1002 W.\ Green Street, Urbana, IL 61801 USA}
\altaffiltext{11}{Department of Physics, University of Illinois Urbana-Champaign, 1110 W.\ Green Street, Urbana, IL 61801 USA}
\altaffiltext{12}{Center for Theoretical Physics and Department of Physics, Massachusetts Institute of Technology, Cambridge, MA 02139}
\altaffiltext{13}{Department of Astronomy, University of California, Berkeley, CA 94720-3411, USA}
\altaffiltext{14}{Las Cumbres Observatory Global Telescope Network, 6740 Cortona Dr., Suite 102, Goleta, CA 93117, USA}
\altaffiltext{15}{Department of Physics, University of California, Santa Barbara, Broida Hall, Mail Code 9530, Santa Barbara,CA 93106, USA}
\altaffiltext{16}{Carnegie Observatories, Las Campanas Observatory, Colina El Pino, Casilla 601, Chile}
\altaffiltext{17}{Department of Physics and Astronomy, Aarhus University, Ny Munkegade 120, DK-8000 Aarhus C, Denmark}
\altaffiltext{18}{Department of Astronomy, Kyoto University, Kitashirakawa-Oiwake-cho, Sakyo-ku, Kyoto 606-8502, Japan}
\altaffiltext{19}{Kavli Institute for the Physics and Mathematics of the Universe (WPI), University of Tokyo, 5-1-5 Kashiwanoha, Kashiwa, Chiba 277-8583, Japan}
\altaffiltext{20}{Department of Astrophysical Sciences, Peyton Hall, Princeton University, Princeton, NJ 08544, USA}
\altaffiltext{21}{Department of Physics \& Astronomy, Western Washington University, 516 High Street, Bellingham, WA 98225}

%%%%%%%%%%%%%%%%%%%%%%%%%%%%%%%%
\begin{abstract}
We report evidence for excess blue light from the Type Ia supernova SN~2012cg at fifteen and sixteen days before maximum $B-$band brightness. The emission is consistent with predictions for the impact of the supernova on a non-degenerate binary companion.  This is the first evidence for emission from a companion to a \emph{normal} SN~Ia.  Sixteen days before maximum light, the $B-V$ color of SN~2012cg is 0.2 mag bluer than for other normal SN~Ia.  At later times, this supernova has a typical SN~Ia light curve, with extinction-corrected $M_B = -19.62 \pm 0.02$ mag and $\Delta m_{15}(B) = 0.86 \pm 0.02$.  Our data set is extensive, with photometry in 7 filters from 5 independent sources.  Early spectra also show the effects of blue light, and high-velocity features are observed at early times.  Near maximum, the spectra are normal with a silicon velocity $v_{Si} = -10,500$ km s$^{-1}$.  Comparing the early data with models by \citet{Kasen10} favors a main-sequence companion of about 6 solar masses.  It is possible that many other SN Ia have main-sequence companions that have eluded detection because the emission from the impact is fleeting and faint.  
\end{abstract}

%%%%%%%%%%%%%%%%%%%%%%%%%%%%%%%%
\section{Introduction}
It is widely accepted that Type Ia supernovae (SN~Ia) are the thermonuclear explosions of carbon-oxygen white dwarfs, and many of them appear to explode near the Chandrasekhar mass \citep[$M_{Ch}$; e.g.][]{Hillebrandt00}, though they may arise from progenitors of other masses as well \citep[e.g.][]{Scalzo14}.
Two general progenitor scenarios are commonly considered for a white dwarf to accrete sufficient mass to approach the Chandrasekhar limit.  In the {\it single degenerate} (SD) model, a non-degenerate binary companion star deposits matter onto a white dwarf.  As the white dwarf nears the Chandrasekhar mass, a thermonuclear runaway is initiated \citep{Whelan73,Nomoto82}.  The {\it double degenerate} (DD) scenario postulates that two carbon-oxygen white dwarfs will merge via gravitational inspiral and explode by subsequent carbon ignition \citep{Webbink84,Iben84}.  Other models are being explored, such as one in which the triggering mechanism of the SN~Ia explosion is the head-on collision of two white dwarfs in a 3-body system \citep{Kushnir13} and the {\it core degenerate} scenario that involves the merger of a white dwarf and the hot core of a massive asymptotic giant branch star \citep{Kashi11}.

There is some evidence that both the SD and DD scenarios contribute to the SN~Ia population \citep[see][for a recent review]{Maoz14}.  For example, {\it Hubble Space Telescope} deep pre-explosion imaging of the site of SN~2011fe rules out evolved companion stars with $M >3.5 M_{\odot}$ \citep{Li2011,Graur14}.  This result does not rule out the SD scenario, even for this individual case, but it does cut a swath through the allowable parameter space.  On the other hand, the SN~Ia PTF11kx had clear signs of interaction with shells of circumstellar medium (CSM).  The details suggest a SD system with a Red Giant companion star in a symbiotic nova configuration \citep{Dilday12}.

Many studies have searched for clues indicating the interaction between a normal SN~Ia and CSM, presumably pointing to the SD scenario, but they are rarely as decisive as PTF11kx.  Detection of interaction based on variable \nai\ features \citep[e.g.][]{Patat07,Blondin09,Sternberg11,Foley12:hires,Maguire13}, and measurements of high velocity features (HVF) from multiple spectral lines \citep[e.g.][]{Marion13,Silverman15} have all been used to investigate
potential SD systems.  See \citet{Maoz14} for a more thorough discussion of work in this area.

A clearer signpost of the SD scenario lies in the very early light curves (LC).  A nearby non-degenerate binary companion will encounter the shock wave and the expanding debris from the explosion. The interaction compresses and heats matter at the point of impact, but not all of the thermal energy is emitted in a prompt burst.  Deeper layers of the ejecta continue to impact the companion and raise the local temperature.  The extra luminosity from this interaction will be strongest in the ultraviolet and blue optical bands, and it will only be detectable for a few days after explosion \citep[][but see \cite{Maeda14,Kutsuna15}]{Kasen10}.

The effect on the observed brightness depends on the viewing angle \citep{Kasen10,Brown12a}.  Interaction is unlikely to be detected if the impact location is too far from the direct line-of-sight to the SN.  Thus, not all explosions in single degenerate systems are expected to produce an early light curve signal; perhaps only $\sim$10\% of cases will do so \citep{Kasen10}.

No previous detections have been reported for the interaction between a \emph{normal} SN~Ia and its companion after inspections of hundreds of LC from SN~Ia, though many of these were not obtained early enough to test the \citet{Kasen10} model predictions.  

\citet{Hayden10} looked for interaction signals in $B-$band LC of 108 SDSS SN~Ia.  They found that companion stars would have to be less than about $6 M_{\odot}$ on the main-sequence and they strongly disfavored Red Giant companions.  \citet{Ganeshalingam11} also examined $B-$band rise time behavior for 61 SN~Ia and found no evidence of companion interaction.  \citet{Tucker11} analyzed $U-$band light curves of $\approx 700$ SN~Ia from the ESSENCE Project and other sources.  They found no signature of shock heating from Red Giant companions.  \citet{Bianco11} determined that less than 10\% out of 87 SN~Ia could have come from white dwarf-Red Giant binary systems.   \citet{Olling15} used high cadence data from $Kepler$ and they found no evidence for interaction in 3 SN~Ia.  The $Kepler$ bandpass is not sensitive below 400nm, and it is not clear if it would detect a shocked companion.

Individual SN~Ia that are found very nearby or very early provide high quality data that may be unavailable to larger surveys.  Analyses of such SN~Ia have revealed an interesting amount of diversity, but no clear signs of shock interaction with a companion.  

\citet{Brown09,Brown12a} reported early UV LC from a few SN~Ia without discovering evidence for excess luminosity.  \citet{Foley12} described excess UV flux from SN~2009ig at early phases, but concluded that the colors were inconsistent with an interaction.  \citet{Margutti12} with X-rays and \citet{Chomiuk12} with radio observations ruled out most of the parameter space for a main-sequence or evolved companion to SN~2011fe.   \citet{Schaefer12} determined that SNR 0509-67.5 in the Large Magellanic Cloud contains no candidates for the companion star to deep limits.  They claim to eliminate all previously published single-degenerate models for this SN~Ia.  \citet{Zheng13} showed that measurements of SN~2013dy, obtained only a few hours after the estimated time of the explosion, do not reveal any evidence for interaction.   \citet{Goobar14} find that a very early section of the SN~2014J LC ($\sim$0.5-2.0 days after explosion) is flatter than a $t^2$ rise; but they rule out the SD model due to constraints on the size of the companion.  \citet{Margutti14} use X-ray non-detections of SN~2014J to rule out single-degenerate systems with steady mass loss.  

\citet{Cao15} report the detection of a significant UV pulse in early data of iPTF14atg.  While intriguing, these results may not be directly applicable to the progenitor scenarios of normal SN~Ia.  This object is part of a class that is very rare \citep{Ganeshalingam12}; it is about 3 magnitudes subluminous compared to a normal SN~Ia, and it does not follow the Phillips relation \citep{Phillips93}.  The current work focuses on the normal SN~Ia, SN 2012cg. This object peaks in $M_B$ between $-$19.4 and $-$19.8 mag and does follow the Phillips relation.  Here we show that the early UV and optical light curves for SN 2012cg are well matched by single degenerate models in which excess blue light is produced by the impact of the supernova on its companion.  Photometric and spectroscopic observations of SN~2012cg and the data reduction details are described in Section~\ref{data}.    Excess luminosity in early photometry and spectra of SN~2012cg is described and analyzed in Section~\ref{excess}.  Predictions of theoretical models for the interaction of a SN~Ia and a binary companion are described in Section~\ref{models} and the models are compared to observations of SN~2012cg.  Section~\ref{spectra} discusses the spectra and the evolution of features in pre-maximum spectra.  Discussion and conclusions are presented in Section~\ref{conc}.

%%%%%%%%%%%%%%%%%%%%%%%%%%%%%%%%
\section{Data Acquisition and Reduction}
\label{data}

SN 2012cg was discovered in Virgo Cluster Galaxy NGC~4424 on May 17.2, 2012 UT = MJD 56065.2 by the Lick Observatory Supernova Search \citep[LOSS;][]{Filippenko01} with the 0.76m Katzman Automatic Imaging Telescope (KAIT).  The discovery was promptly announced to the community by email and by ATEL \citep{Cenko12}.   This rapid notice allowed many observers to begin following SN~2012cg on May 18 which was less than 3 days after the explosion and more than 16 days before the time of maximum brightness in the $B-$band (\bmax).   We report photometric and spectroscopic observations of SN~Ia~2012cg obtained from May 18.2, 2012 (UT) which is 16.1 days before \bmax\ ($-$16.1d) to June 26.0 (+22.7d).

%**************************** Figure 1 ********************************
\begin{figure}[t]
\center
\includegraphics[width=0.5\textwidth]{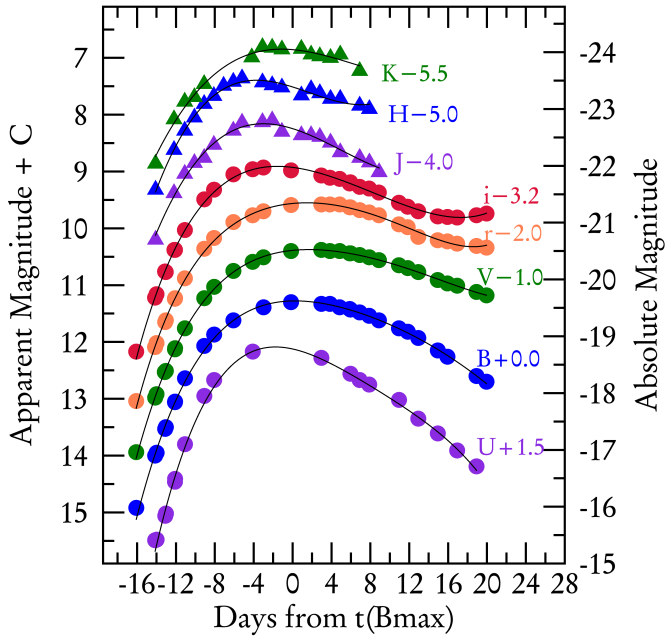}
\caption{$UBVr'i'JHK_s$ photometry of SN~2012cg obtained at the F. L. Whipple Observatory from $-$16.1d to +19.9d.  The data are corrected for MW and host galaxy extinction. The black lines are polynomial fits used to identify the peak brightness and the time of peak for each filter (Table~\ref{peaktbl}).  We find \bmax\ = 560681.3 (MJD) = June 3.3 (UT), with a peak magnitude in the $B-$band of $-$19.62 mag and $\Delta m_{15}(B)$ = 0.86.  \label{plc}}
\end{figure}

%**************************** Figure 2 ********************************
\begin{figure}[t]
\center
\includegraphics[width=0.48\textwidth]{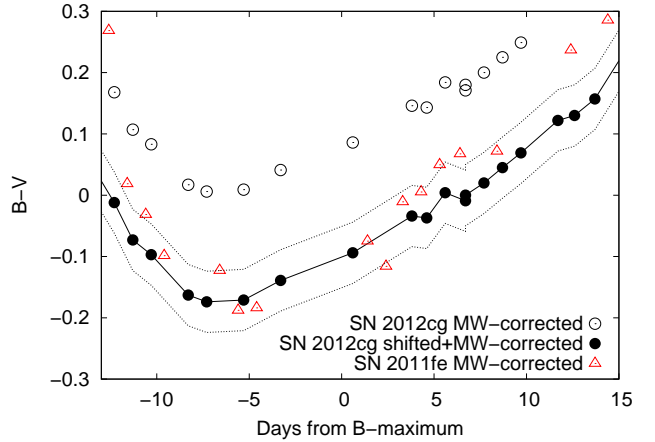}
\caption{$B-V$ color evolution for SN~2012cg and SN~2011fe after correction for Milky Way (Galactic) reddening.  The curves are fit to minimize the difference near \bmax.  We assume that SN~2011fe has essentially zero reddening, so the magnitude of the difference determines $E(B-V)_{host}\approx 0.18$ mag. (See Section~\ref{data}.) \label{bvfig1}}
\end{figure}

%+++++++++++++++++++++++++++++++++++++++++++++++
\subsection{Photometric Observations}

Figure~\ref{plc} shows optical and near infrared (NIR) photometry of SN 2012cg obtained from $-$16.1d to +19.9d. The optical data are from the F. L. Whipple Observatory (FLWO) using the 1.2m telescope and KeplerCam, while the NIR data were obtained with the Peters Automated Infrared Imaging Telescope (PAIRITEL). 

The data in the figure have been corrected for Milky Way (MW) and host extinction.  Polynomial fits give the dates of peak brightness in each filter, the maximum apparent magnitudes and the decline rate parameters, $\Delta m_{15}$ (Table~\ref{peaktbl}).  The absolute magnitudes were computed using $d=15.2 \pm1.9$ Mpc ($\mu=30.90 \pm0.3mag$; Tully-Fisher) for NGC~4424 \citep{Cortes08}.   Uncertainties in the absolute magnitude estimates do not affect our primary results.

The FLWO optical data ($u',B,V,r',i'$) were reduced using IRAF and IDL procedures described in \citet{Hicken07}.  Table~\ref{flwotbl} provides the original measurements without dereddening.  The FLWO galaxy templates were obtained on January 10, 2014, which is 596 days after \bmax.

Transformation to the standard photometric system was performed using local comparison stars around the SN in the same field-of-view. The linear transformation equations were calibrated using \citet{Landolt92} standards for \emph{UBV} and \citet{Smith02} standards for \emph{r'}- and \emph{i'}-bands.  The zero-points of the transformations were determined with data from photometric nights.  The zero-points for images obtained on non-photometric nights were determined by differential aperture photometry (\emph{DAOPHOT}) using tertiary standard stars in the vicinity of the SN.  Further details of these methods can be found in \citet{Hicken12}.

The FLWO \emph{u'}-band can be correlated with \citet{Landolt92} $U-$band magnitudes via the equation $u' = U + 0.854$ mag \citep{Chonis08}.  Table~\ref{flwotbl} shows the measured values for $u'$ while the figures use the values corrected for extinction and converted to $U-$band.
  
NIR images were obtained at the FLWO in the $J,H,K_{s}$ bands by PAIRITEL (Table~\ref{pteltbl}).  The data are processed into mosaics using the PAIRITEL Mosaic Pipeline version 3.6 implemented in python. Photometry is performed on the mosaicked images with DoPHOT \citep{Schechter93} using a modified version of the ESSENCE project photometry pipeline \citep{Miknaitis07}.  Photometric zero points are computed using the 2MASS point source catalog \citep{Cutri03}. Details of PAIRITEL observations and reduction of NIR supernova data can be found in \citet{Friedman12} and \citet{Friedman15}.

We are fortunate to be able to include pre-maximum photometry of SN~2012cg that was obtained at other facilities.  After their discovery of SN~2012cg, LOSS/KAIT continued to monitor the SN and their pre-maximum, uncorrected data are listed in Table~\ref{kaittbl}.  These data were reduced with an image-reduction pipeline described in \citet{Ganeshalingam10}.  The acquisition and reduction of LOSS/KAIT data is described in detail by \citet{Silverman12}.   

KAIT templates were obtained on December 9, 2013 which is 554 days after \bmax.  The galaxy-subtracted photometry produced the same values as reported by \citet{Silverman12} (Zheng, private communication). 

$B-$ and $V-$band photometry was obtained by the Las Cumbres Observatory Global Telescope Network of 1-m telescopes (LCOGT), and reduced using a custom pipeline developed by LCOGT which is based on standard procedures, including \emph{pyraf, DAOPHOT} and \emph{SWARP} in a python framework. Instrumental magnitudes were transformed to the standard system \citep{Landolt92} using standard star observations obtained on photometric nights \citep{Brown13}.  The pre-maximum $B-$ and $V-$band data from the LCOGT are listed in Tabel~\ref{lcogttbl}.

\emph{Swift} photometric data were obtained with the Ultraviolet/Optical Telescope \citep[UVOT;][]{Roming05}.  The UVOT reduction used a pipeline developed for the Swift Optical/Ultraviolet Supernova Archive (SOUSA; \citealp{Brown14}).  It is based on \citet{Brown09}, including subtraction of the host galaxy count rates and uses the revised UV zeropoints and time-dependent sensitivity from \citet{Breeveld11}.  Table~\ref{swifttbl1} shows the $v, b, u, uvw1, uvm2$, and $uvw2$ measurements of SN~2012cg from $-$15.7d to +1.0d.  The \emph{Swift} galaxy template was obtained by observing the location of SN~2012cg 428 days after \bmax.  
 
ROTSE-IIIb uses an unfiltered CCD and an automated image differencing analysis to search for SNe~\citep{Yuan08}.  Photometry is calibrated to an effective $r-$band magnitude by comparing to USNO B1.0~\citep{usno}.   For comparison with other data sources in this paper, the ROTSE clear data, which include significant $B-$ and $U-$band sensitivity, have their zero-point adjusted to B-band.   ROTSE detected SN2012cg on May 17.178, which is 1.1 hrs before the discovery epoch reported by \citet{Silverman12} (May 17.223). Prediscovery ROTSE images on May 16.177 yielded no detection to a limiting magnitude of 16.9.  Table~\ref{rotsetbl} gives the ROTSE measurements.

The ROTSE data are reduced by differential aperture photometry using IDL procedures adapted from DAOPHOT, followed by subtraction of the underlying host.  Transformation to Bessel V-band was performed using reference stars to a radius of 3' from SN 2012cg.  The transformation was calibrated by comparison to the APASS catalog.  This measurement is compared to image differencing where it exists.  There is good agreement with the host subtraction result except for the first epoch when the SN is dim and image differencing is difficult.  Additional uncertainties at all epochs are extracted from the measured variation due to altering the aperture photometry parameters.

%+++++++++++++++++++++++++++++++++++++++++++++++
\subsection{Reddening}
\label{reddening}
To estimate reddening due to the host galaxy, we compared the $B-V$ color curves of SN~2012cg to SN~2011fe after applying $E(B-V)_{MW} = 0.018$ mag for the Milky Way extinction in the direction of SN 2012cg \citep{Schlafly11}.   $B-$ and $V-$band LC of SN~2011fe were obtained from the Piszk\'estet{\H o} Mountain Station of the Konkoly Observatory, Hungary.  These data were previously published by \citet{Vinko12}. 

The color curves are aligned at \bmax\ by applying a vertical shift of $E(B-V) = 0.18$ mag.  SN~2011fe is essentially unreddened by its host \citep{Nugent11,Vinko12,Chomiuk13}, so this difference provides a plausible estimate of the reddening of SN~2012cg from the host galaxy.  It is the same $B-V$ offset found by \citet{Silverman12} for SN~2012cg.   Figure~\ref{bvfig1} illustrates the comparison between the two de-reddened color curves. The dotted lines mark the $\pm 0.05$ mag uncertainty of $E(B-V)_{host}$ for SN~2012cg.   

Summing the Milky Way and host galaxy components of the extinction, we determine that the total reddening of SN~2012cg is $E(B-V)_{total} = 0.18 + 0.018 = 0.198 \approx 0.20 \pm 0.05$ mag, which is consistent with that reported by \citet{Silverman12}.  

We also use the hierarchical Bayesian statistical model BayeSN to fit the $BVr'i'JH$ light curve data from FWLO and PAIRITEL.  BayeSN models the variations in observed optical and NIR SN Ia light curves as a combination of an intrinsic light curve distribution and a distribution of host galaxy dust extinction, determined from a low-z SN Ia training set \citep{Mandel11}. Applied to an individual SN Ia, it computes the posterior probability of its light curve and dust parameters.  Using this method, for SN~2012cg, we
inferred a host galaxy dust color excess of $E(B-V)_{host} = 0.25 \pm 0.03$ (assuming $R_V = 3.1$).  

Differences in the total reddening on the order of 0.08 mag have much smaller effect on the B-V curve of SN 2012cg than the color excess we measure (see \S \ref{excess}). In \S \ref{colors} we show that the color excess for SN~2012cg with respect to other SNe Ia is about 0.4 - 0.5 mag, which is significantly larger than the uncertainty of E(B-V) given above.  We can increase the uncertainties to $\pm 0.08$ mag include the BayeSN result and the \citet{Amanullah15} result of $E(B-V)_{total} = 0.15 \pm 0.02$ mag without affecting our primary results.

%+++++++++++++++++++++++++++++++++++++++++++++++
\subsection{Estimating \bmax\ and $t=0$}
\label{txp}
We estimate \bmax\ by fitting a polynomial to the $B-$band LC (Figure~\ref{plc}).  We find \bmax\ = MJD 56081.3 $\pm$ 0.5d (June 3.3 UT), with $M_B = -19.62\pm0.08$ mag(Table~\ref{peaktbl}).    The phases of all reported observations of SN~2012cg are defined as relative to the time of \bmax.  Our estimate for \bmax\ is between the estimates provided by \citet{Silverman12} (June 2.0 UT) and  \citet{Munari13} (June 5.0 UT). 

We also estimate the rise time from explosion ($t=0$) to \bmax\ of SN~2012cg to compare the timing of observations with other SN.  This estimate has no effect on our measurements of excess flux.  The $L \propto t^2$ ``fireball" model for early light curves of SN~Ia \citep{Arnett82} produces a good fit to the data for many well observed SN \citep{Nugent11}.  The \citet{Arnett82} model derives $L \propto t^2$ from solving a diffusion equation, but the result is much like the simple assumption that with a constant temperature and expansion velocity, flux scales with the surface area which is proportional to $t^2$.  We note however, that recent studies have shown that the LC for some SN~Ia are not well fit by the $t^2$ model \citep[eg.][]{Piro13}. Note also that the $t^2$ model is strictly valid only for the bolometric light curve. Although  $t^2$ works for optical band passes as well in some circumstances, in general for filtered LCs one may expect deviations from the strict $n=2$ exponent. It seems reasonable to use a more general $t^n$ model having different $n$ indices for different filter bands. We find the $n$ does vary among the bands in our fits (\S \ref{excess}). 

\citet{Zheng13} fit the LC SN~2013dy with a variable, or ``broken", power law that has a very steep rise (a higher power law index) for the first day.  At a little more than one day after first light, this model adopts a more gradual curve with an exponent of $2.24 \pm0.08$.  \citet{Zheng14} find that a similar model is required to fit the LC of SN~2014J.  \citet{Dessart14} present LC for delayed detonation and pulsational-delayed detonation models of SN~Ia that do not fit well to the $t^2$ model.  

Despite differences in their trajectories back toward $t=0$, if we ignore the breakout phase, then all of these models are monotonically decreasing without inflections in the model LC.  At this time, it is not obvious that one model is preferred to another, so we use the classic $t^2$ model to find the moment of explosion.  The model LC is fit to the FLWO data between $-$12d and $-$8d which produces an estimated time for the explosion of MJD=56062.5 $\pm$ 0.5d (May 15.5 UT).  This result is consistent with \citet{Silverman12} who estimate the explosion date to be May 15.7 (UT).  

These results give a rise time of 18.8 days.  A variable power law model \citep{Zheng13,Zheng14} would estimate an explosion date of about 1 day later and a rise time about 1 day faster.  \citet{Silverman12} measure an earlier time of \bmax\ (June 2.0 UT), so their rise time estimate is shorter at 17.3 days. 

%+++++++++++++++++++++++++++++++++++++++++++++++
\subsection{Zero-Point Corrections}
\label{zp}
To calibrate and compare the early-time data of SN~2012cg from different sources, a $(t-t_{exp})^2$ model LC was fit to the $V-$ and $B-$band data .  The model was selected by minimizing the residuals for the FLWO data at phases from $-$14d to $-$10d.  Each of the other data sets was fit to the $t^2$ curve by moving it up or down to minimize the residuals of the data from each source over the same phase interval.  Thus the shape of each LC was preserved while systematic errors in the photometric calibration of the LC from the various sources was reduced.  

The zero-point offsets used for all the light curve figures presented herein are:  0.00 mag for KAIT $V-$band; $-0.08$ mag for KAIT $B-$band; $-0.12$ mag for LCOGT $V-$band; $-0.24$ mag for LCOGT $B-$band; $-0.16$ mag for $Swift$ $V-$band; $-0.08$ mag for $Swift$ $B-$band.  The ROTSE data, has its zero-point adjusted to $B-$band.

We note that the $-$17d KAIT data in both $V-$ and $B-$bands (blue triangles) were measured from data near the detection limit of the instrument; the error bars come from the formal measurement uncertainty and are thus likely underestimated \citep{Silverman12}.  

 %                            V             B     \\
 %FLWO (CfA) &     0.00 &     0.00 \\
 %KAIT (UCB) &      0.00 & $-$0.08 \\
 %LCOGT        & $-$0.12 & $-$0.24 \\
 %$Swift$         & $-$0.16 & $-$0.08 \\ 

%**************************** Figure 3 ********************************
\begin{figure}[t]
\center
\includegraphics[width=0.5\textwidth]{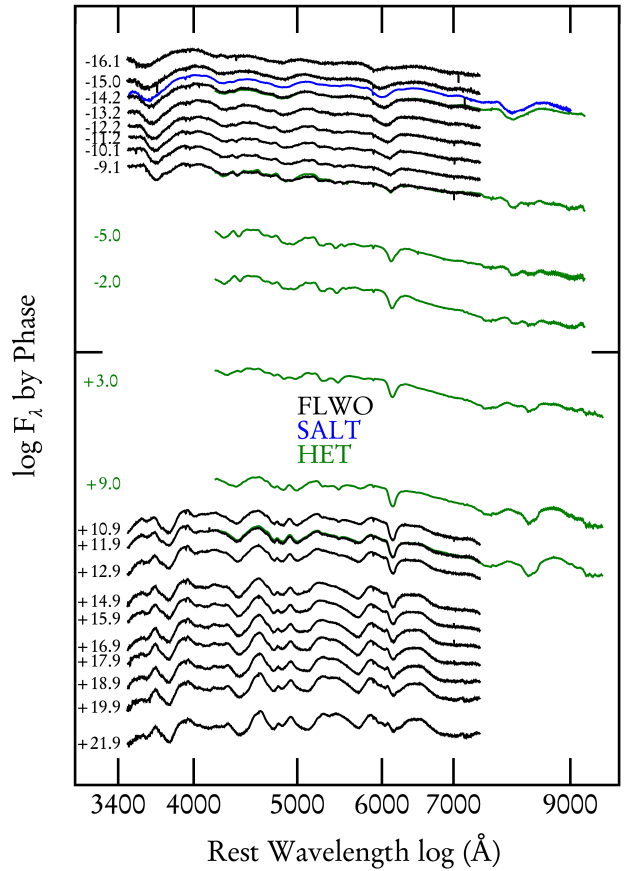}
\caption{Optical spectroscopy of SN~2012cg obtained from $-$16.1d to +21.9d.  The sources are FLWO (black), SALT (blue) and HET (green).  The continuum slopes of the spectra show that SN~2012cg is a blue SN~Ia at early phases.  Absorption features in the pre-maximum spectra are relatively weak.  This is characteristic of slightly overluminous SN~Ia.   \label{ospec}}
\end{figure}

%+++++++++++++++++++++++++++++++++++++++++++++++
\subsection{Spectroscopic Observations}

Optical spectra of SN~2012cg were obtained from May 18.2 through June 25.2 (Figure~\ref{ospec}).  These dates correspond to phases $-$16.1d to +21.9d.  Details for the observations are in Table~\ref{opttbl}.

Optical spectra (3480--7420~\AA, displayed in black) were obtained with the FLWO 1.5m Tillinghast telescope and the FAST spectrograph \citep[FAST;][]{Fabricant98}.  The position angle was 90 degrees but the airmass was low ($\le$ 1.18).  FAST data are reduced using a combination of standard IRAF and custom IDL procedures \citep{Matheson05}.  Additional optical spectra (4,200$-$10,100~\AA, green) were obtained with the 9.2m Hobby-Eberly Telescope \citep[HET;][]{Ramsey98} at the McDonald Observatory using the Marcario Low-Resolution Spectrograph \citep[LRS;][]{Hill98}.  HET/LRS spectra are reduced with standard IRAF procedures.  The HET spectra of SN~2012cg obtained on May 20, 25 and 29 were previously published by \citet{Silverman12}.  

One early epoch spectrum was obtained with Robert Stobie Spectrograph (RSS) on the Southern African Large Telescope (SALT) covering the range 3500$-$9000 \AA\ (blue). This was reduced with a custom pipeline that incorporates routines from PyRAF and PySALT \citep{Crawford10}.

Optical spectra from the FLWO, HET and SALT cover slightly different wavelength ranges.  None of the sources provide continuous coverage through these phases (see Table~\ref{opttbl}). When they do overlap ($-$14d, $-$9d, 12d), the agreement is excellent.

%**************************** Figure  4 *****************************
\begin{figure*}[t]
\center
\includegraphics[width=0.88\textwidth]{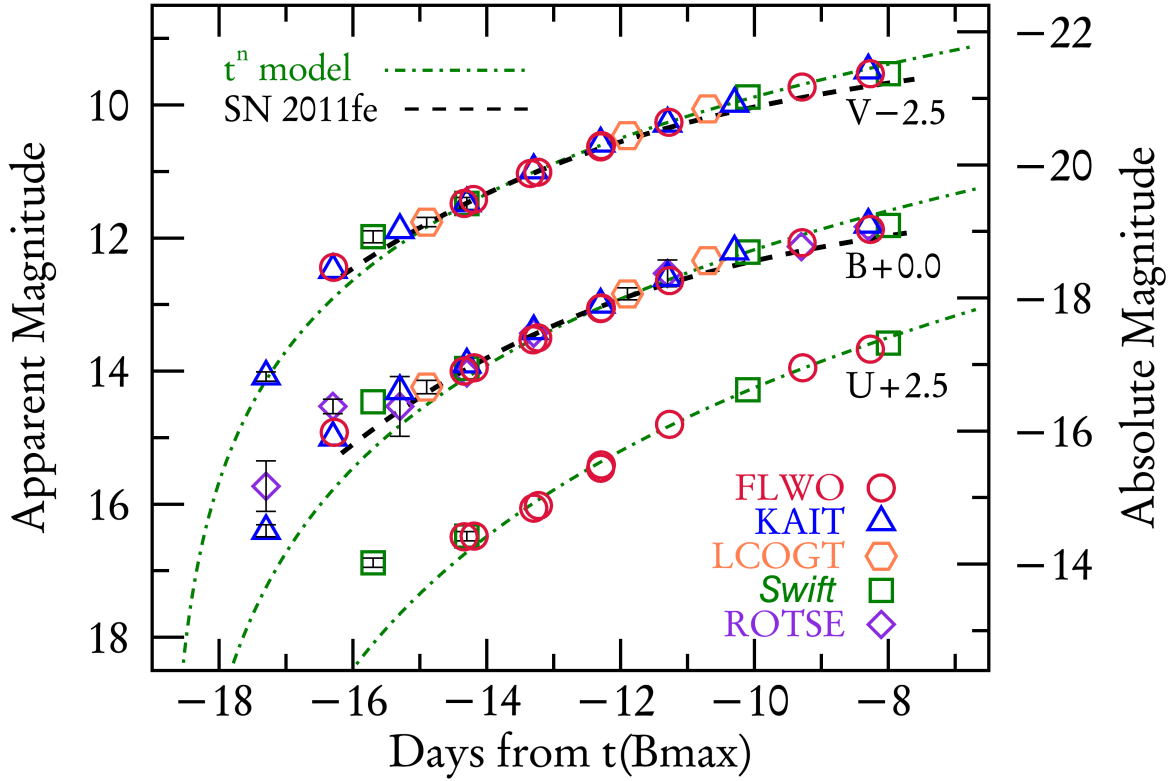}
\caption{$V-$, $B-$ and $U-$band photometry (top to bottom) of SN~2012cg from multiple sources. The data have been corrected for extinction. The LC of SN~2011fe \citep[dashed black lines]{Vinko12} stretched to yield a rise time of 18.8 days equivalent to SN~2012cg (see text) and $t^n$ model LCs  (dot-dash, green) are plotted as templates for a normal SN~Ia. The power-law indices used for different bands are 3.4 ($U$), 2.4 ($B$) and 2.2 ($V$), see text for details. In all bands, the SN~2012cg data display excess flux at phases earlier than $-$14d. From $-$14d toward maximum light, the data from SN~2012cg and SN~2011fe fit the templates well. Uncertainties are marked only where they approach the size of the symbols. Note that uncertainties on the earliest KAIT points are likely underestimated (see \S \ref{models} for more information).  \label{pxrn}}
\end{figure*}

%**************************** Figure  5 ********************************
\begin{figure}[t]
\center
\includegraphics[width=0.5\textwidth]{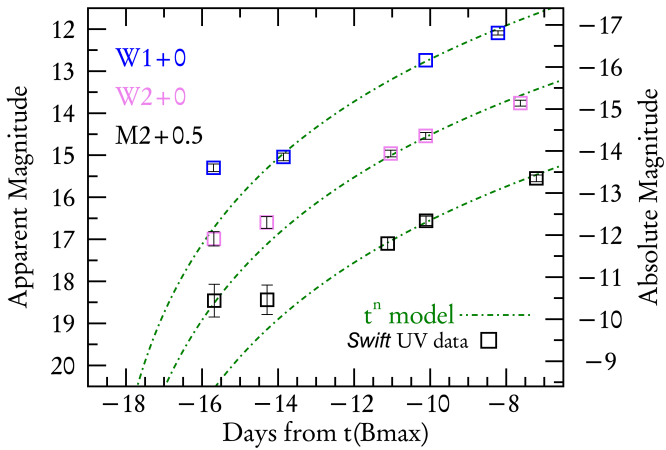}
\caption{$Swift$ photometry (squares) for SN~2012cg obtained in UV filters: $W1, M2$ and $W2$.  The data have been corrected for extinction.  Excess flux is apparent at $-$16d in all filters, and it is also present at $-$14d for $M2$ and $W2$.   Model LCs scaled as $t^{3.6}$ (see text) are plotted for reference in each passband with green dot-dashed lines.  The timing of the observed UV excess matches well with the optical data displayed in Figure~\ref{pxrn}.  \label{pxsw}}
\end{figure}

%%%%%%%%%%%%%%%%%%%%%%%%%%%%%%%%
\section{Detection of Excess Luminosity}
\label{excess}

Figures~\ref{pxrn} and \ref{pxsw} display early photometry in six filters from five sources that reveal excess luminosity in the very early LC of SN~2012cg. These data have been corrected for extinction using $E(B-V)_{total} = 0.20$ mag (see \S \ref{reddening}) and the \citet{fitzp99} reddening law with $R_V = 3.1$. By ``excess" we mean luminosity measurements that exceed the fit model function by at least three times the measurement uncertainty, i.e. they deviate from the models by more than $3 \sigma$. For each passband in Figures~\ref{pxrn}  and \ref{pxsw}, a $t^n$ model light curve (green dot-dashed curves) was fit to the measurements between $-14$d and $-8d$. Regardless of whether $n$ was fixed at 2 or allowed to float, for phases earlier than $-14$d the observed data from various sources consistently exceed the model LC by several times the measurement uncertainty. 

By using strict $t^2$ models we find that between $-18$d and $-14$d the average flux excess ratio, $(f_{obs} - f_{model})/\sigma_{obs}$, is 13.2, 8.6 and 8.8 for the $U$-, $B$- and $V$-band, respectively (note that the amount of excess flux is higher in the $B$-band than in the $V$-band, but the higher uncertainty of the $B$-band observations make their excess flux ratio similar to that of the $V$-band). Using the more general $t^n$ model and letting $n$ float, $n=3.4 \pm 0.1$ ($U$), $n=2.44 \pm 0.05$ ($B$) and $n=2.20 \pm 0.03$ ($V$) are found when fitting the data between $-14$d and $-8$d as before.  In this case the excess flux ratios are somewhat reduced, but their average values are still 9.3 in $U$, 6.6 in $B$ and 6.0 in $V$. Thus, regardless of which models are used for comparison, the earliest observed fluxes deviate from the models by more than $5\sigma$ in the $V$-band and blueward. Note that a model LC that uses a variable power-law with the date of explosion scaled to one day later than the $t^2$ model and higher indices for the first day (Zheng et al. 2013, 2014) would increase the measured excess fluxes even more.

We emphasize that we do not use these simple power-law LC models either to derive physical constraints for the amount of excess flux or to investigate their origin. The physics of the origin of the early-phase excess fluxes will be studied by comparing the data with real physical models in \S \ref{models}. Here we use the $t^n$ models  only for illustration purpose, i.e. to reveal that, unlike other well-observed SN~ Ia, SN~2012cg show noticeable deviations from the simple ``fireball" model
in the blue- and UV-bands.  

Figure~\ref{pxrn} shows measurements from $V-$, $B-$ and $U-$band filters.  The abscissa is time in days from \bmax, and the ordinate is apparent magnitude with absolute magnitude (using $\mu$ = 30.90) displayed on the right axis.  Uncertainties are smaller than the size of the symbols, excepted where indicated.  The optical data were obtained at the FLWO, KAIT, LCOGT, \emph{Swift} and ROTSE.  

SN~2011fe has been described as a good example of a typical SN~Ia \citep[e.g.][]{Nugent11,Chomiuk13}.  We use data of SN~2011fe as a surrogate for a ``normal" SN~Ia, using the well constrained LC parameters of \citet{Vinko12} and \citet{Pereira13}.  In Figure~\ref{pxrn}, $V-$ and $B-$band LC of SN~2011fe \citep{Vinko12} are plotted as black dashed lines.  The extinction-corrected data for SN~2011fe are fit with a polynomial and plotted as a continuous LC in order to avoid crowding the figure with more symbols. 
The rise time for SN~2011fe is 17.7d, whereas our estimate for the rise time of SN~2012cg is 18.8d.  Since we make all direct comparisons based on the phase with respect to \bmax, we stretch the rise time of SN~2011fe to 18.8d for plotting.  

Figure~\ref{pxsw} shows $Swift$ measurements for UV filters: $W1, M2$, and $W2$.   Error bars in the figure and uncertainties in the table take into account the leaks of redder light into both the $W1$ and $W2$ filters.  deleted text It is clear that the UV measurements at $-$16d are well above the power-law model LCs. Using power-law indices similar to those found above for the $U$-band ($n \sim 3.6$) the excess fluxes for the {\it Swift} UV-bands are similar to that in the $U$-band. The larger error bars in the $Swift$ data decrease the excess flux ratios, but they are still significant: 9.1 in $W1$, 3.0 in $M 2$ and 4.9 in $W2$ filters, respectively.  The amount of excess flux is greater in the UV at $-16$d than it is for the $B$ and $V$ optical bands shown in Figure \ref{pxrn}. This result is consistent with model predictions for increased UV luminosity from interaction events (see \S \ref{models}).

%**************************** Figure  6 ********************************
\begin{figure}[t]
\center
\includegraphics[width=0.50\textwidth]{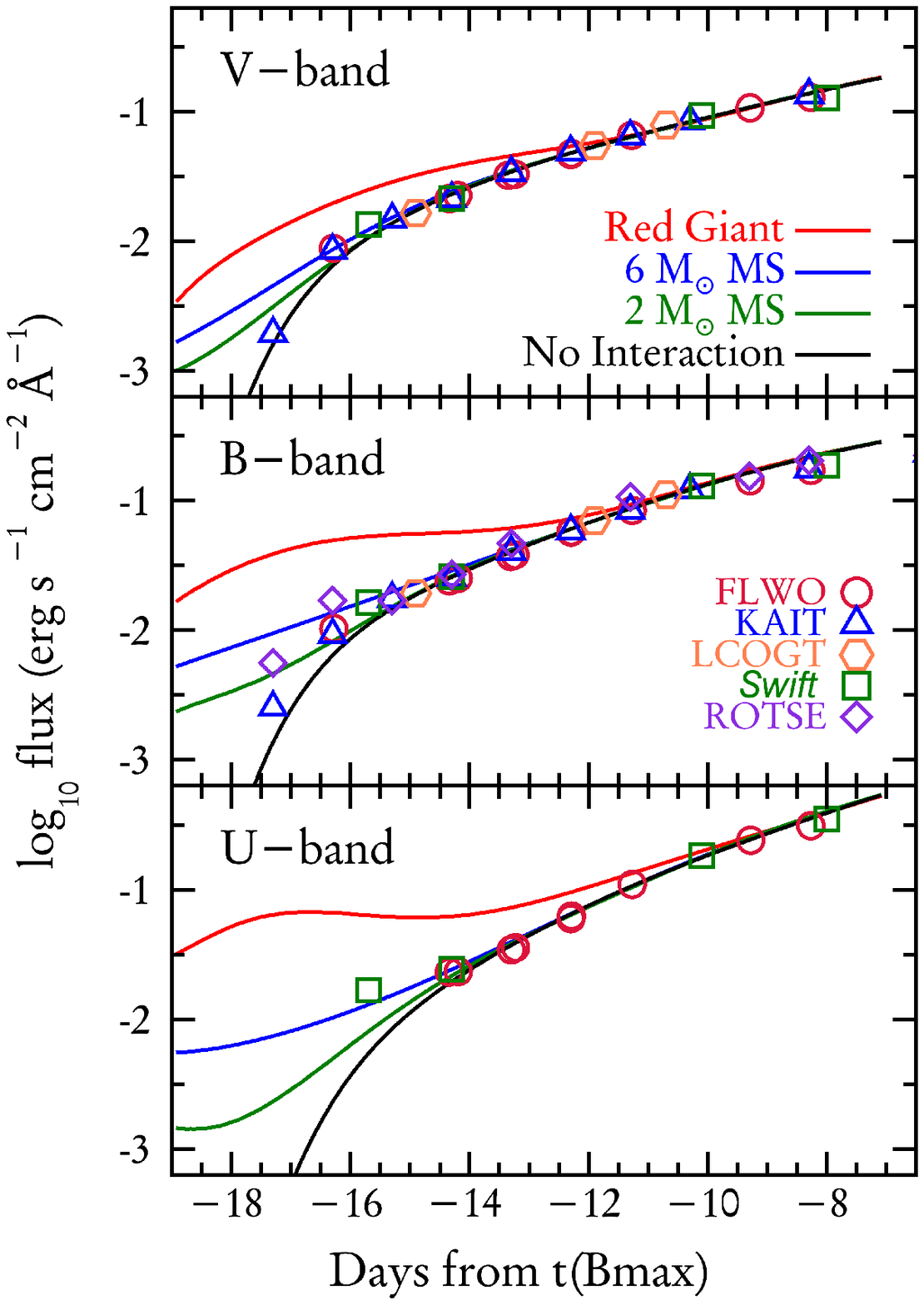}
\caption{Early $U-$, $B-$ and $V-$band data for SN~2012cg are plotted with model results for a normal SN~Ia with no interaction and 3 cases of interaction with a binary companion \citep{Kasen10}.  The models fit the data well at phases after $-$14d (except for the Red Giant model (red line) that is divergent through about $-$11d).  When the model light curves separate, the model for a SN Ia with no interaction (black) has the lowest predicted flux.  Data obtained before $-$14d are brighter than predicted by the normal models and the closest model fit is for interaction with a 6 $M_{\sun}$ MS star (blue line). The values of the earliest KAIT points (blue triangles) are somewhat uncertain (see \S \ref{models} for more information.  \label{pmod}}
\end{figure}

%**************************** Figure 7 ********************************
\begin{figure*}[t]
\centering
\includegraphics[width=0.7\textwidth]{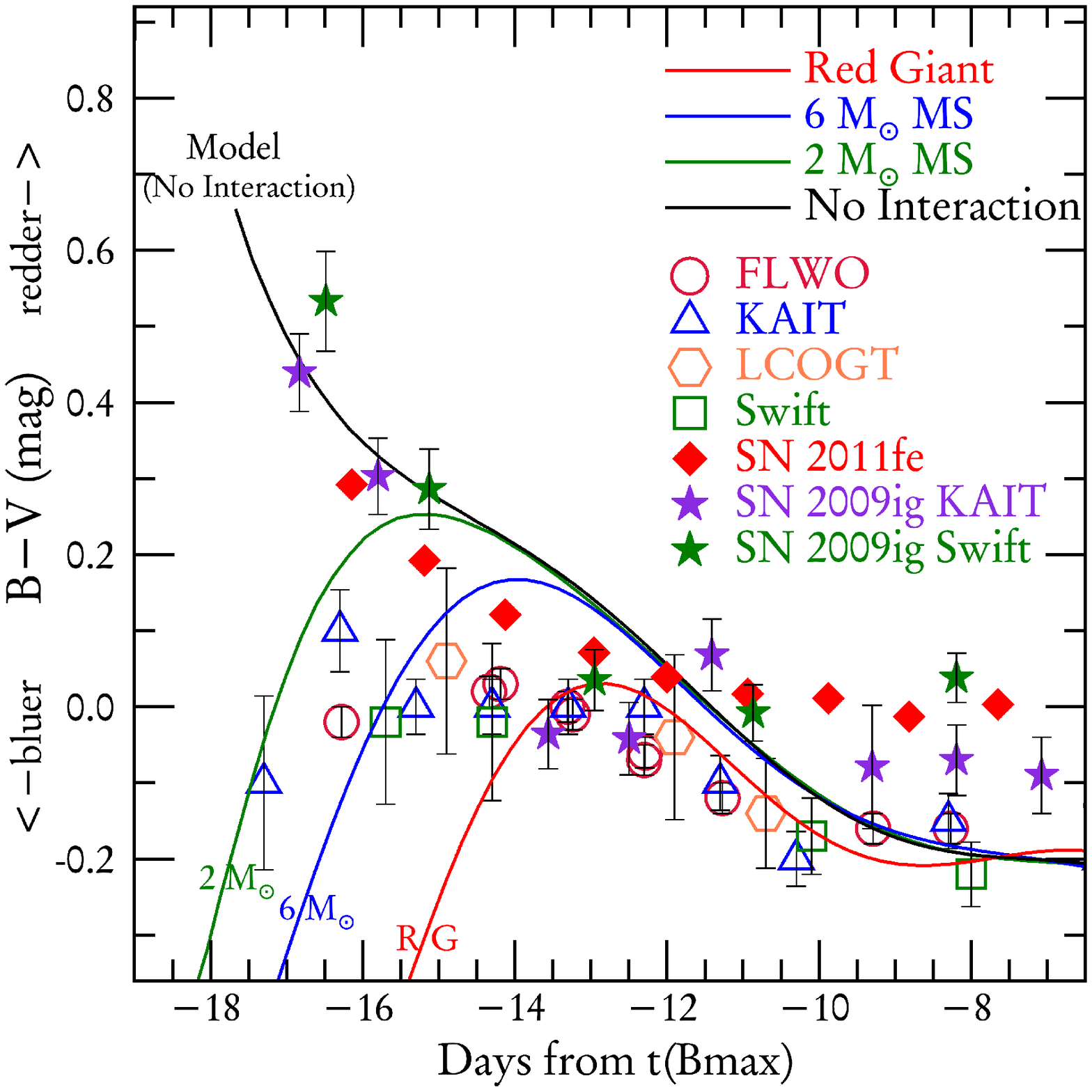}
\caption{$B-V$ colors for the SN~2012cg are displayed using open symbols.  The colored lines trace $B-V$ for the interaction models.  $B-V$ data for two normal SN~Ia are plotted using filled symbols: SNe~2009ig \citep[stars]{Foley13} and 2011fe \citep[diamonds]{Pereira13}.  The colors for the other two SN are similar to the model for a normal SN~Ia with no interaction (solid black line) but they do not fit the data for SN~2012cg or any of the interaction models.  The colors for SN~2012cg are much bluer at the earliest phases, they redden for $2-3$ days and then ``turn over'' near $-$15d.  After reaching this peak, the SN~2012cg colors follow the normal track.  The $B-V$ data for SN~2012cg are closest to the model with a 6 $M_{\odot}$ companion (blue line).    \label{pbv}}
\end{figure*}

%%%%%%%%%%%%%%%%%%%%%%%%%%%%%%%%
\section{Comparing SN~2012\lowercase{cg} to Models}
\label{models}

In the single degenerate model, assuming the absence of circumstellar material, a SN~Ia will expand freely after the explosion until it encounters the companion star.  At the point of impact, matter is compressed and heated while the SN continues to expand.  Material flowing around the companion star forms a bow shock, and a cavity is opened in the SN ejecta as it is diverted around the companion.  Emission from the shock heated region can escape through this hole in the expanding SN \citep{Marietta00,Kasen10,Cao15}.  The size of the cavity is determined by the radius of the companion and that determines the amount of excess radiation that will emerge within the first few hours.     

The outer layers of the expanding material fill in the hole, but the supernova ejecta continue to collide with the binary companion.  These impacting layers are heated by compression and some of the kinetic energy is also dissipated at this point.  Additional thermal energy can diffuse out in the hours and days that follow the initial prompt burst.  Continuing radiative diffusion from deeper layers of ejecta can produce emission in the optical and UV that may exceed the radioactively-powered luminosity of the supernova for a few days after the explosion. 

\citet{Kasen10} modeled the shock from a SN~Ia as it impacts a companion star.  He calculated the observational consequences for three different binary companions: a Red Giant with $r = 2 \times 10^{13}$~cm, a $6 M_{\sun}$ MS star with $r = 2 \times10^{12}$~cm and a $2 M_{\sun}$ MS star with $r = 5 \times 10^{11}$~cm.  The models predict that under optimal conditions, optical and UV emission from the interaction will produce a detectable contribution to the light curves for a few days after the explosion.    After this time, the shock heated emission will no longer contribute to the observed LC, and the SN will behave like a normal SN~Ia with the LC entirely powered by radioactive decay of $^{56}$Ni.

The reddest filter modeled by \citet{Kasen10} is $V$.  Consequently we do not consider the $r'/R$ and $i'/I$ data when discussing the interaction models.  The models were originally defined in AB magnitudes and the observations were measured in the Vega system.  In order to compare them, we changed the magnitude measurements for both data and the models to flux units.  In that format, the models can be moved up or down while preserving the relative flux at all wavelengths.  

In Figure~\ref{pmod}, we show the same $U-$, $B-$ and $V-$band data of SN~2012cg that appear in Figure~\ref{pxrn}.  The data are plotted with model LC from for a normal SN~Ia with no interaction (solid black line) and LC for interactions with 3 possible companions: a Red Giant star (RG, red), a $6 M_{\sun}$ main-sequence star (MS, blue) and a $2 M_{\sun}$ MS star (green) \citet{Kasen10}.  The models for different passbands show that the signatures of interaction have a greater deviation from normal at bluer wavelengths.

With the exception of the RG model that diverges from the other models at later phases, the interaction models remain close together and fit the data well from $-$14d to $-$8d.  Earlier than $-$14d however, the predicted LC for the different models diverge and the separation is sufficient to easily differentiate the model predictions.  

In Figure~\ref{pmod}, the $-$16d and $-$15d data are clearly brighter than the non-interaction models.  There is some scatter in the data, but most of the points lie closest to the blue line which is the model for interaction with a $6 M_{\sun}$ MS companion.  The data are inconsistent with models for a normal SN~Ia or a Red Giant companion viewed on axis.

This result defines a $6 M_{\sun}$ MS star as the smallest allowed companion.  If the impact location were not directly along the line-of-sight to the SN, then the observations of SN~2012cg could be produced by a much larger companion, such as a Red Giant.  As the viewing angle becomes more oblique, the observed excess flux would be reduced.

Off-center interaction sites will be more common than directly aligned sites.  \citet{Kasen10} determined that even for the strongest possible signal from an interaction, the collision of a SN~Ia with a Red Giant companion, the increased flux due to shock-heated ejecta would only be detectible about 10\% of the time.  Smaller companions create smaller and weaker interaction signatures that are even more dependent on viewing angle in order to be detected.  \citet{Brown12a} found similar results for a decrease in the observed flux as the position of the interaction is incrementally offset from the direct line of sight to the SN.  They also find that the observed flux as a function of wavelength changes, with the UV becoming relatively brighter as the viewing angle increases.

%+++++++++++++++++++++++++++++++++++++++++++++++
\subsection{$B-V$ Colors}
\label{colors}

Figure~\ref{pbv} shows $B-V$ colors for SN~2012cg from $-$17d to $-$8d.   Blue colors are negative (toward the bottom of the figure) and red colors are positive (toward the top).  The data and models are the same as those presented in the top 2 panels of Figure~\ref{pmod}. We note that the colors presented here match well with those reported by \citet{Silverman12}.

The \citet{Kasen10} model for a normal SN~Ia without interaction (solid black line) shows $B-V$ colors that are significantly different than the $B-V$ colors for models with interaction.  The non-interacting model is very red (top left) soon after the explosion, due to Fe-group line blending that suppresses the $B$-band continuum.  The $B-V$ colors rapidly become bluer as the relative $B$-band flux increases.  For a non-interacting SN~Ia, $B-V$ is monotonically decreasing during this time period and it never produces a color peak.  

The figure also shows $B-V$ colors for two other normal SN~Ia that have very early observations.  The filled stars are $B-V$ for SN~2009ig \citep{Foley12} and the filled diamonds are $B-V$ for SN~2011fe \citep{Pereira13}.  In both cases, the color curves follow the general path of Kasen's non-interacting model.  They become bluer with time and neither of these normal SN~Ia produces a peak in their color curves.  The phases of both SN~Ia have been stretched to match the 18.8 day rise time of SN~2012cg.  Without stretching, the color curves rise even more steeply into the red.

The KAIT data (purple) and $Swift$ data (green) for SN~2009ig are displayed in different colors. These data from \citet{Foley12} have been corrected for systematic differences in their respective instrumental response functions using spectrophotometry \citep[also known as an S-correction; e.g][]{Stritzinger02} that makes them slightly bluer than without the corrections.  In order to avoid introducing additional systematics, we do not apply S-corrections to any of the photometry of SN~2012cg. The effect would be $\approx0.1$ mag at $-$16d and the applied S-correction would make the early colors of SN~2012cg even bluer.  No significant difference is apparent between the $Swift$ colors and colors from ground based sources.

In contrast, the $B-V$ colors for the interaction models start very blue in the earliest phases.  They grow rapidly redder and then approach the color curve of non-interacting SN Ia after a day or two.  The $B-V$ color curves for the models reach a red peak within a few days and then they decline in unison with the color curve for a non-interacting SN~Ia.   

The early $B-V$ colors of SN~2012cg behave like the interaction models.  They are blue in the earliest data and they reach a red peak near $-$15d.  The data have more scatter than the models, but during the rise to peak the $B-V$ data of SN~2012cg are found near the model results for interactions a $6 M_{\sun}$ MS star.  Irrespective of the best model fit, it is clear that SN~2012cg has a $B-V$ color peak, and that soon after the peak, the $B-V$ colors follow the path of a normal SN~Ia.
 
These early photometric data of SN~2012cg are consistent with the interaction models and are clearly unlike SN~2009ig, SN~2011fe, or the model for a normal SN~Ia that experiences no interaction.  The model for a larger companion, such as the Red Giant, does not fit the timing of the data peak.

%**************************** Figure 8 ********************************
\begin{figure}[t]
\center
\includegraphics[width=0.40\textwidth]{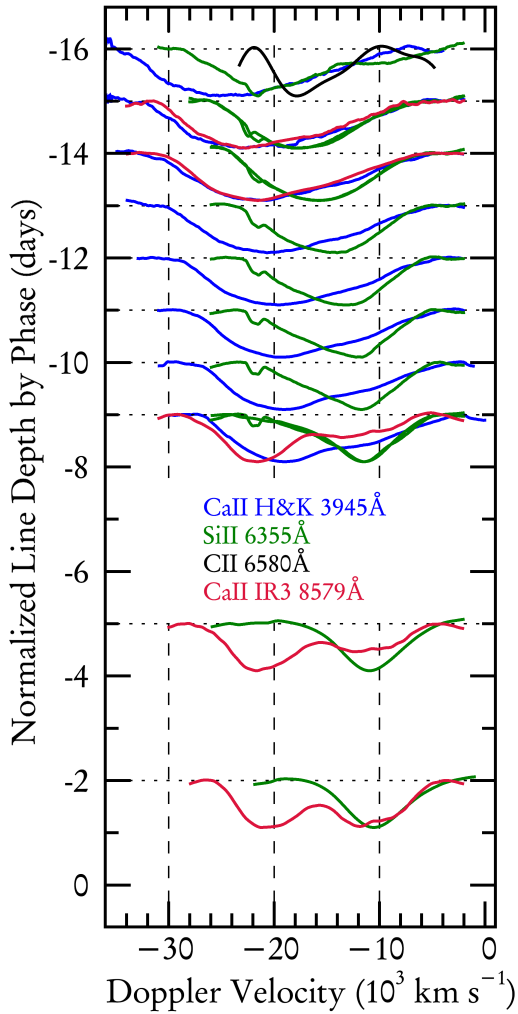}
\caption{The evolution of absorption features \cii\ (black), \si\ (green) and \ca\ (blue and red) in pre-maximum spectra of SN~2012cg.  The features are normalized to a flat continuum and the line depths are normalized to 1.0.  These line profiles identify the locations of line forming regions for each ion in velocity space.  \cii\ is only measurable at $-$16d.  \si\ exhibits a detached HVF at $-$16d and strong influence of the HVF in \si\ line profiles at $-$14d and $-$13d.  HVF of \ca\ are common in SN~Ia, and here they persist until at least $-$2d.  \label{pcompv}}
\end{figure}

%%%%%%%%%%%%%%%%%%%%%%%%%%%%%%%%
\section{Spectra}
\label{spectra}

The spectral features of SN~2012cg are similar to other SN~Ia that are slightly overluminous with moderate decline rates.  All of the typical SN~Ia features are present, but in the earliest spectra, the velocities are slightly higher and features are slightly shallower than features found in spectra from fainter SN~Ia.  By $-$7d, the spectra of SN~2012cg are similar to all normal SN~Ia.

As shown by \citet{Marion13} and \citet{Marion15}, plotting the pre-maximum features of multiple ions in the same velocity space is a productive tool to identify the relative locations of their line forming regions in radial space.  In particular, this technique allows easy identification and comparison of photospheric-velocity features (PVFs) --- absorption features with minima indicating typical SN~Ia photospheric velocities --- and detached, high-velocity features (HVFs) --- absorption features with minima indicating significantly higher velocities than typical SN Ia photospheric velocities \citep[e.g.][]{Marion13,Childress14,Silverman15}. 

Figure~\ref{pcompv} shows the pre-maximum spectra from Figure~\ref{ospec} zoomed in on individual absorption features of  \ca\ H\&K, \si\ \wl 6355, \cii\ \wl 6580 and the \ca\ infrared triplet (IR3; \wl 8579).  The phases of observation are from $-$16d (top) to +0d (bottom).  The features have been normalized to a flat continuum, and the line depths have been normalized to 1.0 in arbitrary flux units.   

The only measurable \cii\ detection is for \wl 6580 at $-$16d.  That relatively narrow absorption feature is plotted in black and can also be seen in the red wing of \si\ \wl 6355 at this phase (near 5,000 \kms).  The \cii\ velocity of about 18,000 \kms\ is clearly lower than the HVF for \si.  \citet{Silverman12} trace \cii\ in SN~2012cg through about $-$8d.  They use SYN++ fitting \citep{Thomas11:synapps}\footnote{https://c3.lbl.gov/es/.}  to tease measurements out of small distortions in the spectra.

\si\ and \ca\ show HVF components that change with time.  \si\ \wl 6355 has a distinct HVF at $-$16d with a velocity of about $-$21,500 \kms\ at the absorption minimum.  From $-$15d to $-$12d, the \si\ HVF and PVF are blended into a single broad absorption feature.  By $-$11d, there is no longer evidence for \si\ HVF.   \si\ PVF are measured to be $-$10,500 \kms\ on $-$9d and the \si\ velocities remain constant through \bmax. 

HVF for \ca\ are present in the spectra of most SN~Ia.  Here, they are prominent from $-$16d to about $-$11d.   HVF \ca\ becomes weaker compared to \ca\ PVF, but they persist through \bmax.   Near $-$10d, the \ca\ HVF feature becomes narrower, the limit of the blue wing moves from about 30,000 \kms\ to about 25,000 \kms\ and it makes a more abrupt transition to the continuum. 

\ca\ PVF are first detected in H\&K beginning about $-$9d.  This phase is also when the primary absorption in \ca\ H\&K (seen in the figure at $-$21,000 \kms) begins to be distorted by \si\ \wl 3858 \citep{Foley13,Marion13}.  The absorption minimum for these features is about $-$11,000 \kms\ in the rest frame of the \si\ line, which suggests that it is indeed PVF \si\ and not HVF \ca.

%++++++++++++++++++++++++++++++++++++++++
\subsection{Early Spectra are Very Blue}

\begin{figure}[t]
\center
\includegraphics[width=0.5\textwidth]{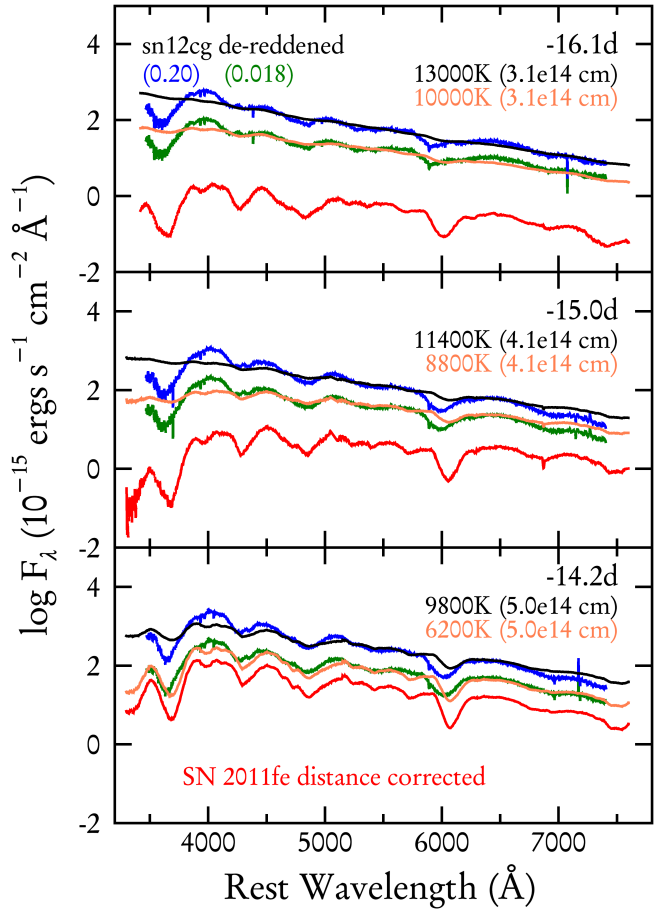}
\caption{The earliest spectra of SN~2012cg are dereddened and plotted in blue for Host $+$ MW extinction and green for MW only.  The spectra plotted in red are distance corrected SN~2011fe spectra from similar phases.  The black and orange spectra are the SN~2011fe data with blackbodies added at the temperatures and radii listed in each panel.  Dilution of the features consistent with adding continuum to the spectra of SN~2011fe.  Note that the BB radii are approximately the same as the photospheric radii at these phases.  The SN~2012cg spectra are bluer and hotter than spectra of SN~2011fe at comparable phases which is consistent with the photometric measurements.  At $-$14.2d, the red spectrum is offset by $-$0.2 log flux units to make it easier to see the orange and green spectra.  \label{scomp}}
\end{figure}

Figure~\ref{scomp} shows the FLWO optical spectra of SN~2012cg obtained at $-$16.1d, $-$15.0d and $-$14.2d.  The spectra are dereddened and plotted in blue when using the combined extinction of the host and MW ($E(B-V)=0.20$), and plotted in green for MW extinction only ($E(B-V)=0.018$).   Reddening details can be found in Section~\ref{data}.  

The SN~2012cg spectra are compared to spectra of SN~2011fe that represent a typical SN~Ia.  They were obtained at $-$16.1d and $-$15.3d \citep{Parrent12} and at $-$14.3d \citep{Pereira13}.  After correction for extinction, the fluxes are scaled to the distance of SN 2012cg and plotted in red.  

The spectra plotted in black are produced by adding the flux of a blackbody (BB) to the spectra from SN~2011fe.  A different BB is added to the spectra from SN~2011fe to form the orange spectra.  The BB parameters are listed in each panel.  The SN~2012cg spectra are bluer and hotter than spectra of SN~2011fe at comparable phases.   This result is consistent with the photometric measurements. 

Fitting BB shapes to the spectra does not generate precise physical measurements  \citep{Kirshner73}.  The BB parameters displayed here are chosen so that when they are added to the SN~2011fe spectra, the resulting continua fit the spectra of SN~2012cg.  The fitting was done by eye, but changes to the temperature or radius of less than 10\% are sufficient to make it obvious that the results do not fit the target spectrum.  

The radii for the BB that produce the closest fits to the SN~2012cg data are similar to the radii for the photosphere in homologous expansion.  For example, we can define a photospheric radius using the time interval from the estimated time of explosion ($-$18.8d) to the time of a observation.  The rates of expansion, estimated from the absorption minima of the Si II 6355 feature (see Figure \ref{pcompv}), are $\sim 16,000$ \kms\ at $-15.0$d and $\sim 14,000$ \kms\ at $-14.2$d. Since the strong HVF component of the Si II 6355 feature prevents reliable measurement of the photospheric component in the $-16.1$d spectrum, we use $\sim 16,000$ \kms\ from the spectrum taken 1 day later ($-15.0$d).

The expansion parameters that determine the photospheric radius corresponding to the earliest spectrum would be:  $t_1=2.7d=2.33 \times 10^5$s and $v_1=16 \times 10^9$~cm s$^{-1}$.  Therefore at a phase of $-$16.1d, we estimate the radius of the photosphere to be: $R_1 \approx 3.7 \times 10^{14}$~cm.  For the phase of $-$15.0d, the velocity stays the same and $R_2 \approx 5.2 \times 10^{14}$~cm.  At $-$14.4d, we apply $v_3 = 14 \times 10^9$~cm s$^{-1}$ to the time interval from $-$15.0d, and find that $R_3 \approx 6.1 \times 10^{14}$~cm.

High temperature blackbodies, as in the top panel, wash out the spectral features of the SN~2011fe spectra.   Lower temperatures, as found in the bottom panel, make only small changes to the features.  The stronger features in SN~2012cg imply that the excess flux is not a pure blackbody. The SN~2012cg features appear to have a higher optical depth at these very early phases than they do at later phases.  This means higher density ejecta in the line forming regions at $-$16d and $-$15d.  By $-$14.4d the spectral differences are small.

%**************************** Figure  9 ********************************

%+++++++++++++++++++++++++++++++++++++++++++++++
\subsection{Narrow Features from the Host}

Significant changes to narrow features from \nai\ D and \ca\ H\&K may be evidence of a process that could produce an observable increase in luminosity. If the SN shock passes through a region of CSM that has a higher density than is usually found around SN~Ia, then \nai\ and \ca\ would be ionized.  As the region cools, the atoms recombine and may produce additional luminosity.  The recombination will also increase the absorption strength of the observed \nai\ and \ca\ lines \citep{Patat07, Simon09, Blondin09, Blondin12}.   

These narrow features are strong in all spectra in this sample, but the resolution of the FAST spectra ($R \approx 2700$) is not intended for measurements on the order of a few hundred kilometers per second.   The pseudo equivalent widths of the \ca\ and \nai\ lines fluctuate within the expected uncertainties due to random noise.  There is no evidence that the line profiles or equivalent widths are any different during the phases at which we measure excess flux ($-$16d and $-$15d) than they are at any other phase in our sample.  Thus, we rule out the possibility that the observed luminosity enhancement is due to interaction with circumstellar material.

%%%%%%%%%%%%%%%%%%%%%%%%%%%%%%%%
\section{Discussion and Conclusions}
\label{conc}

We find excess luminosity at $-$16d and $-$15d in the light curves of SN~Ia~2012cg. The excess is present in data from multiple filters obtained at multiple sources.  The $B-V$ color curves for SN~2012cg are very blue at these phases, and they clearly diverge from $B-V$ models and data of normal SN~Ia.  Spectral evidence is also used to confirm that this short period of excess luminosity, just a few days after the explosion, is real.   

The basic parameters for SN~2012cg describe a slightly overluminous SN~Ia with a moderately slow decline rate and a normal \si\ velocity: $M_B = -19.62$ mag, $\Delta m_{15}(B) = 0.86$ mag and $v_{Si} = -10,500$ \kms\ at \bmax.  These parameters agree with \citet{Silverman12} who were unable to detect the extra flux because their earliest observations were near the detection limit.  

We also examine optical and infrared spectra from SN~2012cg that reveal a blue continuum and relatively weak absorption features in the pre-maximum spectra.  HVF are detected for both \si\ and \ca.  Silicon velocities are normal ($v_{Si} = -10,500$ \kms) at $-$10d and they stay constant through \bmax.

\citet{Kasen10} described how the impact of the SN on a companion could produce emission that may be detected as enhanced luminosity in the first few days after the explosion.  We compare the early LC data and $B-V$ colors of SN~2012cg to the \citet{Kasen10} models for a normal SN~Ia and to models for interaction with 3 different, non-degenerate companions.  We find that the observations of SN~2012cg are consistent with models for the interaction between a SN~Ia and a main-sequence binary companion of about 6 $M_{\sun}$.  

Interaction with a larger companion star is a possibility with the constraints of our data, if the impact site were significantly off-axis with respect to the line of sight to the SN.  The interaction would still produce excess flux but at a reduced level that may be comparable to the observations \citep{Kasen10,Brown12a}.

Our size estimate for the companion is supported by \citet{Graur15}.  They use pre-explosion $HST$ Wide-Field Planetary Camera 2 images to estimate the upper limits on the luminosity of a possible companion to SN~2012cg.  The limits they derive suggest that the brightest possible companion would be either a Red Giant or a $\approx 7 M_{\odot}$ main-sequence star.  The late-time $HST$ WFC3 photometry from \citet{Graur15} is also consistent with this model.  The pre-explosion limits suggest that the progenitor system of SN~2012cg did not have a helium star donor \citep[e.g. ][]{Wang14, Liu10}.  

We note that \citet{Dessart14} present pulsational-delayed detonation (PDD) models of SN~Ia, and some of these models exhibit color peaks similar to the \citet{Kasen10} models.  The PDD models also have increased luminosity at early times, but they do not reach the levels of excess that we observe in SN~2012cg. Comparing the B-V colors of SN~2012cg, SN~2011fe and SN~2009ig, as plotted in Figure  \ref{pbv}, to the results from synthetic photometry of various PDD 
models\footnote{https://www-n.oca.eu/supernova/snia/snia{\_}ddc{\_}pddel.html} 
by Dessart et al. (2014) we find that the PDD model colors are consistent with the observations at $-14$d and afterward, while before this epoch the PDD models predict $B-V > 0.2$ mag, similar to SN~2011fe and SN~2009ig but in disagreement with the bluer ($B-V \sim 0$ mag) colors of SN 2012cg. 

Other models predict that extra blue flux at early times could be the result of radioactive $^{56}$Ni on the outside of the exploding WD.  This material could potentially come from the burning of accreted He on the surface WD progenitor.  \citet{Shen14} discuss He detonation models for single and double degenerate progenitors of SN~Ia.  They find that the He detonation would likely be triggered by a minimum mass He shell.  In the case of a low mass shell, He will only burn to Si and Ca, and not produce radioactive Ni.  Although a larger He shell may burn all the way to $^{56}$Ni, \citet{Shen14} predict that the He detonation will occur at lower masses and trigger a SN~Ia well before the He-shell becomes capable of producing $^{56}$Ni.

\citet{Amanullah15} show that SN~2012cg may have time variable reddening that would change the colors with time.  However, this effect is reported in only the $M2-V$ colors and only nearer to \bmax\ than the phases we discuss here.  They note that $B-V$ has minimal evolution and they specifically discount the presence of circumstellar dust close enough to the SN that it might affect the brightness and colors we observe at these very early phases.  Thus, the unusual early-time colors we see are probably not due to CSM interaction, but instead are coming from interaction with the companion.

%\subsection{Local Environment}
\citet{Cortes06} report that NGC~4424 has a strongly disturbed stellar disk, with groups of young blue stars outside the locations of current star formation.  They suggest that the peculiarities of NGC 4424 are the result of an intermediate-mass merger plus ram
pressure stripping.  SN~2012cg went off on the East side of the host, about 17" away from the nucleus. That puts it outside the current H$\alpha$ emission region, but within regions that mix dust, blue-star-complexes and \ion{H}{1} in emission.  SN~2012cg is not in a region of active, ongoing star formation, but it is in a location where there was star formation in the recent past.

\citet{Crowl08} describe fiber spectroscopy of NGC 4424. Their fibers clearly encompass the location of SN~2012cg.  All of the fiber spectra are averaged to obtain a composite spectrum of the galaxy.  They determine that the luminosity weighted stellar population of the composite spectrum is about 2 Gyr and that star formation shut off about 300-500 Myr ago.

The turbulent region around SN~2012cg makes it difficult to be precise about the star formation history.  Given these constraints, the maximum mass of stars currently present in this location is likely to approach ZAMS $\approx 9 M_{\odot}$.  They will have been formed toward the end of the recent is star formation epoch.  Less massive stars are also possible.  Thus, the environment is suitable to establish a progenitor system in the region of SN~2012cg with a WD of about $1 M_{\odot}$ and a MS companion of about $6 M_{\odot}$. 

This work emphasizes the importance of photometric observations of SN~Ia as early as $-$17d to evaluate the possible interaction with a companion and for more advanced analysis of progenitor systems.  \citet{Foley12} identified excess UV flux from SN~2009ig at early phases, but the colors were inconsistent with an interaction.  SN~2011fe was observed early but the data reveal no evidence for interaction.  Nearly all of the LC of other SN~Ia that have been used to interpret the presence or absence of interaction do not include sufficiently early data. However, these reported non-detections have been used often to suggest that the SD model for SN~Ia is no longer viable and that SN~Ia are exclusively produced in DD progenitor systems.  

Observations at phases between $-$17d and $-$15d suggest that SN~2012cg had a MS binary companion of about $6 M_{\odot}$ when it exploded.  Therefore SN~2012cg must have evolved in a binary system in which only the SN~Ia progenitor was degenerate.   

The size estimate for the companion should be regarded as a minimum, since the star could be much larger but seen at a less favorable angle.   The constraints of timing and the size of the companion demonstrate that the interaction with a companion would go undetected in all but a few of the current data sets.  Had the angle of observation been different, the excess luminosity in SN~2012cg might have gone unobserved as well.

%%%%%%%%%%%%%%%%%%%%%%%%%%%%%%%%%%%%%%%%%%%%%%%%%
\acknowledgments

JV is supported by Hungarian OTKA Grant NN$-$107637. JCW, GHM, JMS and the UT supernova group are supported by NSF grant AST$-$1109801.   JMS is also supported by an NSF Astronomy and Astrophysics Postdoctoral Fellowship under award AST$-$1302771.  The CfA Supernova Program is supported by NSF grants AST$-$1211196 and AST$-$156854 to the Harvard College Observatory.  RPK was supported in part by the National Science Foundation under Grant NSF PHY$-$1125915 to the Kavli Institute for Theoretical Physics.  ASF acknowledges support from a NSF Graduate Research Fellowship, a NASA Graduate Research Program Fellowship, and a NSF STS postdoctoral fellowship under award SES$-$1056580.    We acknowledge the work of C. Klein, D.L. Starr, and J.S. Bloom on the PAIRITEL mosaic data reduction pipeline.  Additional support comes from program GO$-$12540, provided by NASA through a grant from the Space Telescope Science Institute, which is operated by the Association of Universities for Research in Astronomy, Inc., under NASA contract NAS5-26555.  ROTSE data analysis was supported by NASA grant NNX10A196H subaward UTA13-000844.

GHM is a visiting Astronomer at the Infrared Telescope Facility, which is operated by the University of Hawaii under Cooperative Agreement no. NNX-08AE38A with the National Aeronautics and Space Administration, Science Mission Directorate, Planetary Astronomy Program.  P. J. Brown and the $Swift$ Optical/Ultraviolet Supernova Archive are supported by NASA's Astrophysics Data Analysis Program through grant NNX13AF35G.  E.~Y.~H. acknowledges the support provided by the National Science Foundation under Grant No. AST-1008343 and by the Danish Agency for Science, Technology and Innovation through a Sapere Aude Level 2 grant.  This research at Rutgers University was supported by NSF CAREER award AST-0847157 to SWJ, and NSF REU grant PHY-1263280 for YC.  The work by K. Maeda is supported by JSPS KAKENHI (26800100) and MEXT WPI Initiative, Japan.  This paper includes data gathered with the 6.5 meter Magellan Telescopes located at the Las Campanas Observatory, Chile.  This work makes use of observations from the LCOGT network.  The authors make frequent use of David Bishop's excellent webpage listing recent supernovae and valuable references associated with them: www.rochesterastronomy.org/snimages/.

{\it Facilities:}  \facility{FLWO:1.5m (FAST)}, \facility{HET (LRS)}, \facility{Swift (UVOT; UV grism)}, \facility{IRTF (SpeX)}, \facility{FLWO:1.2m (KepCam)}, \facility{LCOGT}, \facility{Lick (KAST)}, \facility{ROTSE IIIb}, \facility{SALT (RSS)}

%\clearpage
%%%%%%%%%%%%%%%%%%%%%%%%%%%%%%%%%%%%%%%%%%%%%%%%%

%++++++++++++++++++++++++++++++++++++++++++++++++++++++
\
\clearpage
\LongTables

%**************************** Table 1 ********************************
\begin{deluxetable*}{lcccc}
%\tabletypesize{\scriptsize}
\tablecolumns{5}
\tablewidth{0pc}
\tablecaption{Peak Magnitudes for SN 2012\lowercase{cg} from Extinction-Corrected FLWO Data\label{peaktbl}}
\tablehead{ \colhead{Band}  & \colhead{MJD-56000\tablenotemark{a}}& \colhead{$m_{\lambda}$\tablenotemark{b}}  
                & \colhead{$M_{\lambda}$\tablenotemark{c}}& \colhead{$\Delta m_{15}$\tablenotemark{d}}}
\startdata
KepCam U &   79.6 &  10.59 & -20.30 &   1.51 \\
KepCam B &   81.3 &  11.26 & -19.63 &   0.86 \\
KepCam V &   82.3 &  11.36 & -19.54 &   0.51 \\
KepCam r' &   81.8 &  11.53 & -19.37 &   0.63 \\
KepCam i' &   79.5 &  12.08 & -18.82 &   0.86 \\
PAIRITEL J &   78.3 &  12.16 & -18.74 & \nodata \\
PAIRITEL H &   77.6 &  12.39 & -18.51 & \nodata \\
PAIRITEL K &   80.2 &  12.35 & -18.55 & \nodata 
\enddata
\tablenotetext{a}{$\pm 0.4$ days.}
\tablenotetext{b}{$\pm 0.04$ mag.}
\tablenotetext{c}{$\pm 0.08$ mag.}
\tablenotetext{d}{$\pm 0.04$ mag.}
\end{deluxetable*}

%%%%%%%%%%%%%%%%%%%%%%%%%%%%%%%%%%%%%%%%%%%%%%%%%%

%**************************** Table 2 ********************************
\begin{deluxetable}{crcc}
\tabletypesize{\scriptsize}
\tablecolumns{4}
\tablewidth{0pc}
\tablecaption{FLWO 1.2m Observations of SN 2012\lowercase{cg}\label{flwotbl}}
\tablehead{\colhead{MJD-56000} & \colhead{Phase\tablenotemark{a}} & \colhead{Mag} & \colhead{Err}}
\startdata
\cutinhead{$u'$-band}
   67.16 & $-$14.1 &    15.84 &     0.03 \\
   67.31 & $-$14.0 &    15.83 &     0.02 \\
   68.20 & $-$13.1 &    15.41 &     0.02 \\
   68.27 & $-$13.0 &    15.37 &     0.02 \\
   69.20 & $-$12.1 &    14.81 &     0.02 \\
   69.21 & $-$12.1 &    14.76 &     0.02 \\
   70.23 & $-$11.1 &    14.15 &     0.02 \\
   72.22 & $-$9.1 &    13.30 &     0.02 \\
   73.23 & $-$8.1 &    13.02 &     0.02 \\
   77.20 & $-$4.1 &    12.52 &     0.02 \\
   84.26 & 3.0 &    12.63 &     0.02 \\
   87.25 & 5.9 &    12.91 &     0.02 \\
   88.20 & 6.9 &    13.02 &     0.02 \\
   89.18 & 7.9 &    13.10 &     0.02 \\
   92.25 & 10.9 &    13.37 &     0.02 \\
   94.23 & 12.9 &    13.70 &     0.02 \\
   96.24 & 14.9 &    13.96 &     0.04 \\
   98.23 & 16.9 &    14.26 &     0.03 \\
  100.22 & 18.9 &    15.54 &     0.04 \\
\cutinhead{$B-$band}
   65.22 & $-$16.1 &    15.74 &     0.02 \\
   67.16 & $-$14.1 &    14.82 &     0.02 \\
   67.31 & $-$14.0 &    14.77 &     0.02 \\
   68.19 & $-$13.1 &    14.35 &     0.02 \\
   68.26 & $-$13.0 &    14.32 &     0.02 \\
   69.20 & $-$12.1 &    13.88 &     0.02 \\
   69.21 & $-$12.1 &    13.87 &     0.02 \\
   70.23 & $-$11.1 &    13.46 &     0.02 \\
   72.21 & $-$9.1 &    12.89 &     0.02 \\
   73.23 & $-$8.1 &    12.69 &     0.02 \\
   75.21 & $-$6.1 &    12.44 &     0.03 \\
   78.31 & $-$3.0 &    12.21 &     0.02 \\
   81.14 & $-$0.2 &    12.12 &     0.02 \\
   84.26 & 3.0 &    12.15 &     0.02 \\
   85.14 & 3.8 &    12.15 &     0.02 \\
   86.14 & 4.8 &    12.21 &     0.02 \\
   87.18 & 5.9 &    12.24 &     0.02 \\
   87.25 & 5.9 &    12.25 &     0.02 \\
   88.20 & 6.9 &    12.30 &     0.02 \\
   89.18 & 7.9 &    12.36 &     0.02 \\
   90.18 & 8.9 &    12.44 &     0.02 \\
   92.24 & 10.9 &    12.58 &     0.03 \\
   93.15 & 11.8 &    12.64 &     0.02 \\
   94.23 & 12.9 &    12.75 &     0.02 \\
   96.24 & 14.9 &    12.97 &     0.03 \\
   97.23 & 15.9 &    13.08 &     0.02 \\
  100.22 & 18.9 &    13.42 &     0.02 \\
  101.25 & 19.9 &    13.52 &     0.02 \\
\cutinhead{$V-$band}
   65.21 & $-$16.1 &    15.56 &     0.01 \\
   67.16 & $-$14.1 &    14.60 &     0.01 \\
   67.30 & $-$14.0 &    14.54 &     0.01 \\
   68.15 & $-$13.2 &    14.15 &     0.01 \\
   68.26 & $-$13.0 &    14.13 &     0.01 \\
   69.20 & $-$12.1 &    13.75 &     0.01 \\
   69.21 & $-$12.1 &    13.73 &     0.01 \\
   70.22 & $-$11.1 &    13.38 &     0.01 \\
   72.21 & $-$9.1 &    12.85 &     0.01 \\
   73.23 & $-$8.1 &    12.65 &     0.01 \\
   75.21 & $-$6.1 &    12.37 &     0.01 \\
   77.19 & $-$4.1 &    12.21 &     0.01 \\
   78.30 & $-$3.0 &    12.13 &     0.03 \\
   81.14 & $-$0.2 &    12.02 &     0.01 \\
   84.25 & 2.9 &    12.00 &     0.01 \\
   85.14 & 3.8 &    12.02 &     0.01 \\
   86.14 & 4.8 &    12.02 &     0.02 \\
   87.17 & 5.9 &    12.06 &     0.01 \\
   87.25 & 5.9 &    12.06 &     0.01 \\
   88.20 & 6.9 &    12.09 &     0.01 \\
   89.18 & 7.9 &    12.13 &     0.01 \\
   90.18 & 8.9 &    12.18 &     0.01 \\
   92.24 & 10.9 &    12.27 &     0.01 \\
   93.15 & 11.8 &    12.32 &     0.01 \\
   94.23 & 12.9 &    12.39 &     0.01 \\
   96.23 & 14.9 &    12.53 &     0.01 \\
   97.23 & 15.9 &    12.59 &     0.01 \\
   98.23 & 16.9 &    12.63 &     0.01 \\
  100.22 & 18.9 &    12.74 &     0.01 \\
  101.25 & 19.9 &    12.80 &     0.01 \\
\cutinhead{$r'-$band}
   65.21 & $-$16.1 &    15.54 &     0.01 \\
   67.15 & $-$14.2 &    14.59 &     0.01 \\
   67.30 & $-$14.0 &    14.52 &     0.01 \\
   68.19 & $-$13.1 &    14.15 &     0.01 \\
   68.26 & $-$13.0 &    14.12 &     0.01 \\
   69.19 & $-$12.1 &    13.74 &     0.01 \\
   69.21 & $-$12.1 &    13.73 &     0.01 \\
   70.22 & $-$11.1 &    13.38 &     0.01 \\
   72.21 & $-$9.1 &    12.86 &     0.01 \\
   73.22 & $-$8.1 &    12.67 &     0.01 \\
   75.21 & $-$6.1 &    12.39 &     0.01 \\
   77.19 & $-$4.1 &    12.27 &     0.01 \\
   78.30 & $-$3.0 &    12.20 &     0.01 \\
   81.14 & $-$0.2 &    12.09 &     0.01 \\
   84.25 & 2.9 &    12.08 &     0.01 \\
   85.14 & 3.8 &    12.08 &     0.01 \\
   86.14 & 4.8 &    12.08 &     0.01 \\
   87.17 & 5.9 &    12.13 &     0.01 \\
   87.24 & 5.9 &    12.13 &     0.01 \\
   88.20 & 6.9 &    12.18 &     0.01 \\
   89.18 & 7.9 &    12.22 &     0.01 \\
   90.18 & 8.9 &    12.27 &     0.01 \\
   92.24 & 10.9 &    12.43 &     0.01 \\
   93.14 & 11.8 &    12.48 &     0.01 \\
   94.23 & 12.9 &    12.65 &     0.09 \\
   96.23 & 14.9 &    12.71 &     0.01 \\
   97.23 & 15.9 &    12.73 &     0.01 \\
   98.23 & 16.9 &    12.77 &     0.01 \\
  100.22 & 18.9 &    12.81 &     0.01 \\
  101.25 & 19.9 &    12.84 &     0.01 \\
\cutinhead{$i'-$band}
   65.21 & $-$16.1 &    15.71 &     0.01 \\
   67.15 & $-$14.2 &    14.76 &     0.01 \\
   67.30 & $-$14.0 &    14.70 &     0.01 \\
   68.18 & $-$13.1 &    14.30 &     0.07 \\
   68.25 & $-$13.1 &    14.31 &     0.01 \\
   69.14 & $-$12.2 &    13.92 &     0.01 \\
   69.20 & $-$12.1 &    13.92 &     0.01 \\
   70.22 & $-$11.1 &    13.57 &     0.01 \\
   72.21 & $-$9.1 &    13.03 &     0.01 \\
   73.22 & $-$8.1 &    12.86 &     0.01 \\
   75.21 & $-$6.1 &    12.59 &     0.01 \\
   77.19 & $-$4.1 &    12.50 &     0.01 \\
   78.30 & $-$3.0 &    12.47 &     0.03 \\
   81.14 & $-$0.2 &    12.52 &     0.01 \\
   84.25 & 2.9 &    12.61 &     0.01 \\
   85.14 & 3.8 &    12.65 &     0.01 \\
   86.14 & 4.8 &    12.67 &     0.01 \\
   87.17 & 5.9 &    12.75 &     0.01 \\
   87.24 & 5.9 &    12.73 &     0.01 \\
   88.20 & 6.9 &    12.81 &     0.01 \\
   89.17 & 7.9 &    12.84 &     0.01 \\
   90.17 & 8.9 &    12.91 &     0.01 \\
   92.24 & 10.9 &    13.09 &     0.01 \\
   93.14 & 11.8 &    13.16 &     0.01 \\
   94.22 & 12.9 &    13.24 &     0.01 \\
   96.23 & 14.9 &    13.33 &     0.01 \\
   97.23 & 15.9 &    13.34 &     0.01 \\
   98.23 & 16.9 &    13.35 &     0.01 \\
  100.22 & 18.9 &    13.31 &     0.01 \\
  101.25 & 19.9 &    13.28 &     0.01 
\enddata
\tablenotetext{a}{Estimated date of \bmax: MJD 56081.3 = June 3.3 (UT).}
\end{deluxetable}

%%%%%%%%%%%%%%%%%%%%%%%%%%%%%%%%%%%%%%%%%%%%%%%%%%

%**************************** Table 3 ********************************
\begin{deluxetable}{crcc}
\tabletypesize{\scriptsize}
\tablecolumns{4}
\tablewidth{0pc}
\tablecaption{PAIRITEL Observations of SN 2012\lowercase{cg}\label{pteltbl}}
\tablehead{\colhead{MJD-56000} & \colhead{Phase\tablenotemark{a}} & \colhead{Mag} & \colhead{Err}}
\startdata
\cutinhead{$J-$band}
    67.2 &    $-$14.1 &    14.36 &     0.05 \\
    69.1 &    $-$12.2 &    13.54 &     0.06 \\
    70.2 &    $-$11.1 &    13.20 &     0.08 \\
    71.2 &    $-$10.1 &    13.01 &     0.02 \\
    72.2 &     $-$9.1 &    12.92 &     0.09 \\
    73.2 &     $-$8.1 &    12.69 &     0.02 \\
    75.2 &     $-$6.1 &    12.44 &     0.02 \\
    76.1 &     $-$5.2 &    12.30 &     0.08 \\
    78.2 &     $-$3.1 &    12.29 &     0.04 \\
    79.2 &     $-$2.1 &    12.26 &     0.08 \\
    80.2 &     $-$1.1 &    12.46 &     0.10 \\
    82.2 &      0.9 &    12.52 &     0.07 \\
    83.2 &      1.9 &    12.52 &     0.02 \\
    84.1 &      2.8 &    12.55 &     0.04 \\
    85.2 &      3.9 &    12.65 &     0.03 \\
    86.2 &      4.9 &    12.82 &     0.06 \\
    88.2 &      6.9 &    12.92 &     0.02 \\
    89.2 &      7.9 &    13.00 &     0.03 \\
    90.2 &      8.9 &    13.17 &     1.13 \\
 \cutinhead{$H-$band}
    67.2 &    $-$14.1 &    14.42 &     0.11 \\
    69.1 &    $-$12.2 &    13.72 &     0.02 \\
    70.2 &    $-$11.1 &    13.38 &     0.02 \\
    71.2 &    $-$10.1 &    13.15 &     0.03 \\
    72.2 &     $-$9.1 &    12.91 &     0.02 \\
    73.2 &     $-$8.1 &    12.77 &     0.02 \\
    74.2 &     $-$7.1 &    12.59 &     0.06 \\
    75.2 &     $-$6.1 &    12.52 &     0.06 \\
    76.1 &     $-$5.2 &    12.46 &     0.08 \\
    78.2 &     $-$3.1 &    12.53 &     0.02 \\
    79.2 &     $-$2.1 &    12.58 &     0.04 \\
    80.2 &     $-$1.1 &    12.62 &     0.05 \\
    82.2 &      0.9 &    12.76 &     0.09 \\
    83.2 &      1.9 &    12.65 &     0.07 \\
    84.1 &      2.8 &    12.72 &     0.03 \\
    85.2 &      3.9 &    12.82 &     0.02 \\
    86.2 &      4.9 &    12.82 &     0.04 \\
    88.2 &      6.9 &    12.94 &     0.02 \\
    89.2 &      7.9 &    13.00 &     1.52 \\
\cutinhead{$Ks$-band}
    67.2 &    $-$14.1 &    14.42 &     0.11 \\
    69.1 &    $-$12.2 &    13.64 &     0.08 \\
    70.2 &    $-$11.1 &    13.33 &     0.09 \\
    71.2 &    $-$10.1 &    13.25 &     0.04 \\
    72.2 &     $-$9.1 &    13.02 &     0.02 \\
    77.1 &     $-$4.2 &    12.55 &     0.07 \\
    78.2 &     $-$3.1 &    12.37 &     0.07 \\
    79.2 &     $-$2.1 &    12.38 &     0.04 \\
    80.2 &     $-$1.1 &    12.41 &     0.02 \\
    82.2 &      0.9 &    12.40 &     0.03 \\
    83.2 &      1.9 &    12.50 &     0.04 \\
    84.1 &      2.8 &    12.53 &     0.03 \\
    85.2 &      3.9 &    12.56 &     0.02 \\
    86.2 &      4.9 &    12.50 &     0.08 \\
    88.2 &      6.9 &    12.79 &     0.10 
\enddata
\tablenotetext{a}{Estimated date of \bmax: MJD 56081.3 = June 3.3.}
\end{deluxetable}

%%%%%%%%%%%%%%%%%%%%%%%%%%%%%%%%%%%%%%%%%%%%%%%%%
%\clearpage
%\vspace{2in}

%**************************** Table 4 ********************************
\begin{deluxetable}{crcc}
\tabletypesize{\scriptsize}
\tablecolumns{4}
\tablewidth{0pc}
\tablecaption{KAIT (Premaximum) Observations of SN 2012\lowercase{cg}\label{kaittbl}}
\tablehead{\colhead{MJD-56000} & \colhead{Phase\tablenotemark{a}} & \colhead{Mag} & \colhead{Err}}
\startdata
 \cutinhead{$B-$band}
    64.2 &    $-$17.1 &    17.28 &     0.09 \\
    65.2 &    $-$16.1 &    15.94 &     0.05 \\
    66.2 &    $-$15.1 &    15.24 &     0.03 \\
    67.2 &    $-$14.1 &    14.83 &     0.03 \\
    68.2 &    $-$13.1 &    14.33 &     0.03 \\
    69.2 &    $-$12.1 &    13.90 &     0.03 \\
    70.2 &    $-$11.1 &    13.50 &     0.03 \\
    71.2 &    $-$10.1 &    13.11 &     0.03 \\
    73.2 &     $-$8.1 &    12.67 &     0.03 \\
    75.2 &     $-$6.1 &    12.36 &     0.03 \\
    76.2 &     $-$5.1 &    12.30 &     0.03 \\
    78.2 &     $-$3.1 &    12.11 &     0.03 \\
    79.2 &     $-$2.1 &    12.09 &     0.03 \\
    81.2 &     $-$0.1 &    12.12 &     0.03 \\
\cutinhead{$V-$band}
    64.2 &    $-$17.1 &    17.15 &     0.07 \\
    65.2 &    $-$16.1 &    15.56 &     0.02 \\
    66.2 &    $-$15.1 &    14.96 &     0.02 \\
    67.2 &    $-$14.1 &    14.58 &     0.02 \\
    68.2 &    $-$13.1 &    14.08 &     0.02 \\
    69.2 &    $-$12.1 &    13.74 &     0.02 \\
    70.2 &    $-$11.1 &    13.36 &     0.02 \\
    71.2 &    $-$10.1 &    13.08 &     0.02 \\
    73.2 &     $-$8.1 &    12.62 &     0.02 \\
    75.2 &     $-$6.1 &    12.36 &     0.02 \\
    76.2 &     $-$5.1 &    12.28 &     0.02 \\
    78.2 &     $-$3.1 &    12.08 &     0.02 \\
    79.2 &     $-$2.1 &    12.12 &     0.02 \\
    81.2 &     $-$0.1 &    12.04 &     0.02 
\enddata
\tablenotetext{a}{Estimated date of \bmax: MJD 56081.3 = June 3.3.}
\end{deluxetable}

%%%%%%%%%%%%%%%%%%%%%%%%%%%%%%%%%%%%%%%%%%%%%%%%%

%**************************** Table 5 ********************************

\begin{deluxetable}{crcc}
\tabletypesize{\scriptsize}
\tablecolumns{4}
\tablewidth{0pc}
\tablecaption{LCOGT (Premaximum) Observations of SN 2012\lowercase{cg}\label{lcogttbl}}
\tablehead{\colhead{MJD-56000} & \colhead{Phase\tablenotemark{a}} & \colhead{Mag} & \colhead{Err}}
\startdata
\cutinhead{$B-$band}
    66.6 &    $-$14.7 &    15.23 &     0.10 \\
    69.6 &    $-$11.7 &    13.87 &     0.09 \\
    70.8 &    $-$10.5 &    13.35 &     0.06 \\
    75.6 &     $-$5.7 &    12.40 &     0.08 \\
    83.6 &      2.3 &    12.02 &     0.05 \\
\cutinhead{$V-$band}
    66.6 &    $-$14.7 &    15.00 &     0.07 \\
    69.6 &    $-$11.7 &    13.72 &     0.06 \\
    70.8 &    $-$10.5 &    13.33 &     0.04 \\
    75.6 &     $-$5.7 &    12.40 &     0.06 \\
    83.6 &      2.3 &    11.97 &     0.05 
\enddata
\tablenotetext{a}{Estimated date of \bmax: MJD 56081.3 = June 3.3.}
\end{deluxetable}

%%%%%%%%%%%%%%%%%%%%%%%%%%%%%%%%%%%%%%%%%%%%%%%%%
%\clearpage

%**************************** Table 6 ********************************

\begin{deluxetable}{crcc}
\tabletypesize{\scriptsize}
\tablecolumns{4}
\tablewidth{0pc}
\tablecaption{$Swift$ (Premaximum) Observations of SN 2012\lowercase{cg}\label{swifttbl1}}
\tablehead{\colhead{MJD-56000} & \colhead{Phase\tablenotemark{a}} & \colhead{Mag} & \colhead{Err}}
\startdata
\cutinhead{$U-$band}
    65.8 &    $-$15.5 &    15.46 &     0.07 \\
    67.2 &    $-$14.1 &    15.13 &     0.07 \\
    71.4 &     $-$9.9 &    12.91 &     0.05 \\
    73.5 &     $-$7.8 &    12.20 &     0.03 \\
    80.7 &     $-$0.6 &    11.72  &     0.03 \\
    82.4 &          1.1 &    11.78 &     0.03 \\
\cutinhead{$B-$band}
    65.8 &    $-$15.5 &    15.48 &     0.06 \\
    67.2 &    $-$14.1 &    14.93 &     0.05 \\
    71.4 &     $-$9.9 &    13.25 &     0.04 \\
    73.5 &     $-$7.8 &    12.77 &     0.04 \\
    80.7 &     $-$0.6 &    12.16 &     0.03 \\
    82.4 &          1.1 &    12.19 &     0.03 \\
\cutinhead{$V-$band}
    65.8 &    $-$15.5 &    15.23 &     0.09 \\
    67.2 &    $-$14.1 &    14.62 &     0.06 \\
    71.4 &     $-$9.9 &    13.09 &     0.04 \\
    73.5 &     $-$7.8 &    12.61 &     0.04 \\
    75.2 &     $-$6.1 &    12.42 &     0.03 \\
    78.9 &     $-$2.4 &    12.14 &     0.03 \\
    80.6 &     $-$0.7 &    12.08 &     0.03 \\
    82.5 &          1.2 &    12.04 &     0.03 \\
\cutinhead{$UVW1$-band}
    65.8 &    $-$15.5 &    17.14 &     0.12 \\
    67.6 &    $-$13.7 &    16.89 &     0.11 \\
    68.4  &   $-$12.9 &   16.46  &     0.18\\
    71.4 &     $-$9.9 &    14.60 &     0.05 \\
    73.3 &     $-$8.0 &    13.94 &     0.04 \\
    75.2 &     $-$6.1 &    13.57 &     0.04 \\
    78.9 &     $-$2.4 &    13.33 &     0.04 \\
    80.6 &     $-$0.7 &    13.39 &     0.04 \\
    82.5 &          1.2 &    13.48 &     0.03 \\
\cutinhead{$UVM2$-band}
    65.8 &    $-$15.5 &    19.82 &     0.35 \\
    67.2 &    $-$14.1 &    19.86 &     0.32 \\
    70.4 &    $-$10.9 &    18.55 &     0.25 \\
    71.4 &     $-$9.9 &    18.02 &     0.12 \\
    74.3 &     $-$7.0 &    17.04 &     0.11 \\
    75.2 &     $-$6.1 &    16.76 &     0.10 \\
    78.9 &     $-$2.4 &    16.50 &     0.10 \\
    80.6 &     $-$0.7 &    16.49 &     0.09 \\
    82.5 &          1.2 &    16.48 &     0.09 \\
\cutinhead{$UVW2$-band}
    65.8 &    $-$15.5 &    18.83 &     0.18 \\
    67.2 &    $-$14.1 &    18.43 &     0.15 \\
    70.5 &    $-$10.8 &    16.79 &     0.11 \\
    71.4 &     $-$9.9 &    16.34 &     0.11 \\ 
    73.9 &     $-$7.4 &    15.73 &     0.10 \\
    74.4  &    $-$6.9  &   15.51 &     0.09 \\
    75.2 &     $-$6.1 &    15.40 &     0.07 \\
    78.9 &     $-$2.4 &    15.14 &     0.08 \\
    80.6 &     $-$0.7 &    15.18 &     0.07 \\
    82.5 &           1.2 &    15.19 &     0.06 \\
\enddata
\tablenotetext{a}{Estimated date of \bmax: MJD 56081.3 = June 3.3.}
\end{deluxetable}

%%%%%%%%%%%%%%%%%%%%%%%%%%%%%%%%%%%%%%%%%%%%%%%%%

%**************************** Table 8 ********************************

\begin{deluxetable}{crcc}
\tabletypesize{\scriptsize}
\tablecolumns{4}
\tablewidth{0pc}
\tablecaption{ROTSE (Premaximum) Observations of SN 2012\lowercase{cg}\label{rotsetbl}}
\tablehead{\colhead{MJD-56000} & \colhead{Phase\tablenotemark{a}} & \colhead{Mag} & \colhead{Err}}
\startdata
\cutinhead{Clear Filter}
    64.2 &    $-$17.1 &    16.63 &     0.38 \\
    65.2 &    $-$16.1 &    15.41 &     0.11 \\
    66.2 &    $-$15.1 &    15.39 &     0.45 \\
    67.2 &    $-$14.1 &    14.88 &     0.03 \\
    68.2 &    $-$13.1 &    14.32 &     0.03 \\
    70.2 &    $-$11.1 &    13.37 &     0.20 \\
    72.2 &     $-$9.1 &    12.95 &     0.03 \\
    73.2 &     $-$8.1 &    12.70 &     0.03 \\
    75.2 &     $-$6.1 &    12.55 &     0.02 \\
    76.2 &     $-$5.1 &    12.43 &     0.02 
\enddata
\tablenotetext{a}{Estimated date of \bmax: MJD 56081.3 = June 3.3.}
\end{deluxetable}

%%%%%%%%%%%%%%%%%%%%%%%%%%%%%%%%%%%%%%%%%%%%%%%%%
\clearpage

%**************************** Table 9 ********************************

\begin{deluxetable*}{rrclc}
\tabletypesize{\scriptsize}
\tablecolumns{5}
\tablewidth{0pc}
\tablecaption{Optical Spectra of SN~2012\lowercase{cg}\label{opttbl}}
\tablehead{\colhead{Date}  & \colhead{Phase\tablenotemark{a}} & \colhead{Telescope/Instrument} & 
                  \colhead{Range (\AA)} & \colhead{R ($\lambda / \Delta\lambda$)}}
\startdata
  May  18.21 & $-$16.1 & FLWO/FAST & 3300--7400 & 2700 \\
  May  19.30 & $-$15.0 & FLWO/FAST & 3300--7400 & 2700 \\
  May  19.81 & $-$14.5 & SALT/RSS    & 3500-9000 & 1100 \\
  May  20.15 & $-$14.2 & FLWO/FAST & 3300--7400 & 2700 \\
  May  20.20 & $-$14.1 & HET/LRS & 4100--10200 & 1500 \\
   May  21.15 & $-$13.2 & FLWO/FAST & 3300--7400 & 2700 \\
  May  22.15 & $-$12.2 & FLWO/FAST & 3300--7400 & 2700 \\
  May  23.15 & $-$11.2 & FLWO/FAST & 3300--7400 & 2700 \\
  May  24.16 &  $-$10.1 & FLWO/FAST & 3300--7400 & 2700 \\
  May  25.16 &  $-$9.1 & FLWO/FAST & 3300--7400 & 2700 \\
  May  25.20 &  $-$9.0 & HET/LRS & 4100--10200 & 1500 \\
  May  29.20 &  $-$5.0 & HET/LRS & 4100--10200 & 1500 \\
  June   1.20 &  $-$2.0 & HET/LRS & 4100--10200 & 1500 \\
  June   6.20 &    +3.0 & HET/LRS & 4100--10200 & 1500 \\
  June  12.20 &   +9.0 & HET/LRS & 4100--10200 & 1500 \\
  June  14.21 & +10.9 & FLWO/FAST & 3300--7400 & 2700 \\
  June  15.15 & +11.9 & FLWO/FAST & 3300--7400 & 2700 \\
  June  15.20 & +12.0 & HET/LRS & 4100--10200 & 1500 \\
  June  16.25 & +12.9 & FLWO/FAST & 3300--7400 & 2700 \\
  June  18.15 & +14.9 & FLWO/FAST & 3300--7400 & 2700 \\
  June  19.19 & +15.9 & FLWO/FAST & 3300--7400 & 2700 \\
  June  20.24 & +16.9 & FLWO/FAST & 3300--7400 & 2700 \\
  June  21.22 & +17.9 & FLWO/FAST & 3300--7400 & 2700 \\
  June  22.15 & +18.9 & FLWO/FAST & 3300--7400 & 2700 \\
  June  23.18 & +19.9 & FLWO/FAST & 3300--7400 & 2700 \\
  June  25.17 & +21.9 & FLWO/FAST & 3300--7400 & 2700 
\enddata
\tablenotetext{a}{Phase in days with respect to the estimated date of \bmax: MJD 56081.3 = June 3.3.}
\end{deluxetable*}

\end{document}